\documentclass[final]{ias2}

\usepackage{graphicx} 
\usepackage{multirow}
\usepackage{array} 

\usepackage{hyperref} 

\begin{document}

\markboth{Overview of the CKM Matrix}{Tim Gershon}

\title{Overview of the CKM Matrix}

\author[warwick,cern]{Tim Gershon} 
\email{T.J.Gershon@warwick.ac.uk}
\address[warwick]{Department of Physics, University of Warwick, Coventry, United Kingdom}
\address[cern]{European Organization for Nuclear Research (CERN), Geneva, Switzerland}

\begin{abstract}
  The current status of the determination of the elements of the Cabibbo-Kobayashi-Maskawa quark-mixing matrix is reviewed.
  Tensions in the global fits are highlighted.
  Particular attention is paid to progress in, and prospects for, measurements of $CP$ violation effects.
\end{abstract}

 
\maketitle


\section{Introduction}

The Cabibbo-Kobayashi-Maskawa (CKM) matrix~\cite{Cabibbo:1963yz,Kobayashi:1973fv} describes the mixing between the three different families of quark in the Standard Model (SM) of particle physics.
It is therefore a $3\times 3$ unitary matrix, and can be written in terms of four real parameters.
For example, in the Wolfenstein parametrisation~\cite{Wolfenstein:1983yz,Buras:1994ec} it can be expressed
\begin{eqnarray}
  \label{eq:ckm}
  V_{\rm CKM} & = & 
  \left(
  \begin{array}{ccc}
    V_{ud} & V_{us} & V_{ub} \\
    V_{cd} & V_{cs} & V_{cb} \\
    V_{td} & V_{ts} & V_{tb} \\
  \end{array}
  \right) \\
  & = &
  \left(
  \begin{array}{ccc}
    1 - \lambda^2/2 & \lambda & A \lambda^3 ( \rho - i \eta ) \\
    - \lambda & 1 - \lambda^2/2 & A \lambda^2 \\
    A \lambda^3 ( 1 - \rho - i \eta ) & - A \lambda^2 & 1 \\
  \end{array}
  \right) + {\cal O}\left( \lambda^4 \right) \, ,
\end{eqnarray}
where the expansion parameter $\lambda$ is the sine of the Cabibbo angle ($\lambda = \sin \theta_{\rm C} \approx V_{us}$).
With four independent parameters, a $3\times 3$ unitary matrix cannot be forced to be real-valued, and hence $CP$ violation arises as a consequence of the fact that the couplings for quarks and antiquarks have different phases, {\it i.e.} $V_{\rm CKM} \neq V_{\rm CKM}^*$.
In the SM, all $CP$ violation in the quark sector arises from this fact, which is encoded in the Wolfenstein parameter $\eta$.
Moreover, all flavour-changing interactions of the quarks are described by the four parameters of the CKM matrix, which makes it a remarkably predictive paradigm, describing phenomena from the lowest energies (such as nuclear transitions and pion decays) to the highest ($W$ boson and top quark decays) within the realm of accelerator-based particle physics.
Inevitably, a very broad range of theoretical tools (such as chiral perturbation theory, lattice quantum chromodynamics, and heavy quark effective theories) is needed to relate such diverse processes to the underlying physical parameters.


Only a brief summary of the current status is given here, with most attention to experimental progress (discussions of progress in theory can be found in Refs.~\cite{dighe,lubicz}).  More detailed reviews can be found, for example, in Refs.~\cite{Nakamura:2010zzi,Asner:2010qj,Antonelli:2009ws,Charles:2004jd,Bona:2005vz}, and in the summaries of CKM2010, the $6^{\rm th}$ International Workshop on the CKM Unitarity Triangle~\cite{Spadaro:2011uw,Laiho:2011uj,Gorbahn:2011pd,Kreps:2011cb,Fleischer:2011ne,Graham:2011dk}.

\begin{figure}[ht]
  \begin{center}
    \includegraphics[width=0.36\columnwidth]{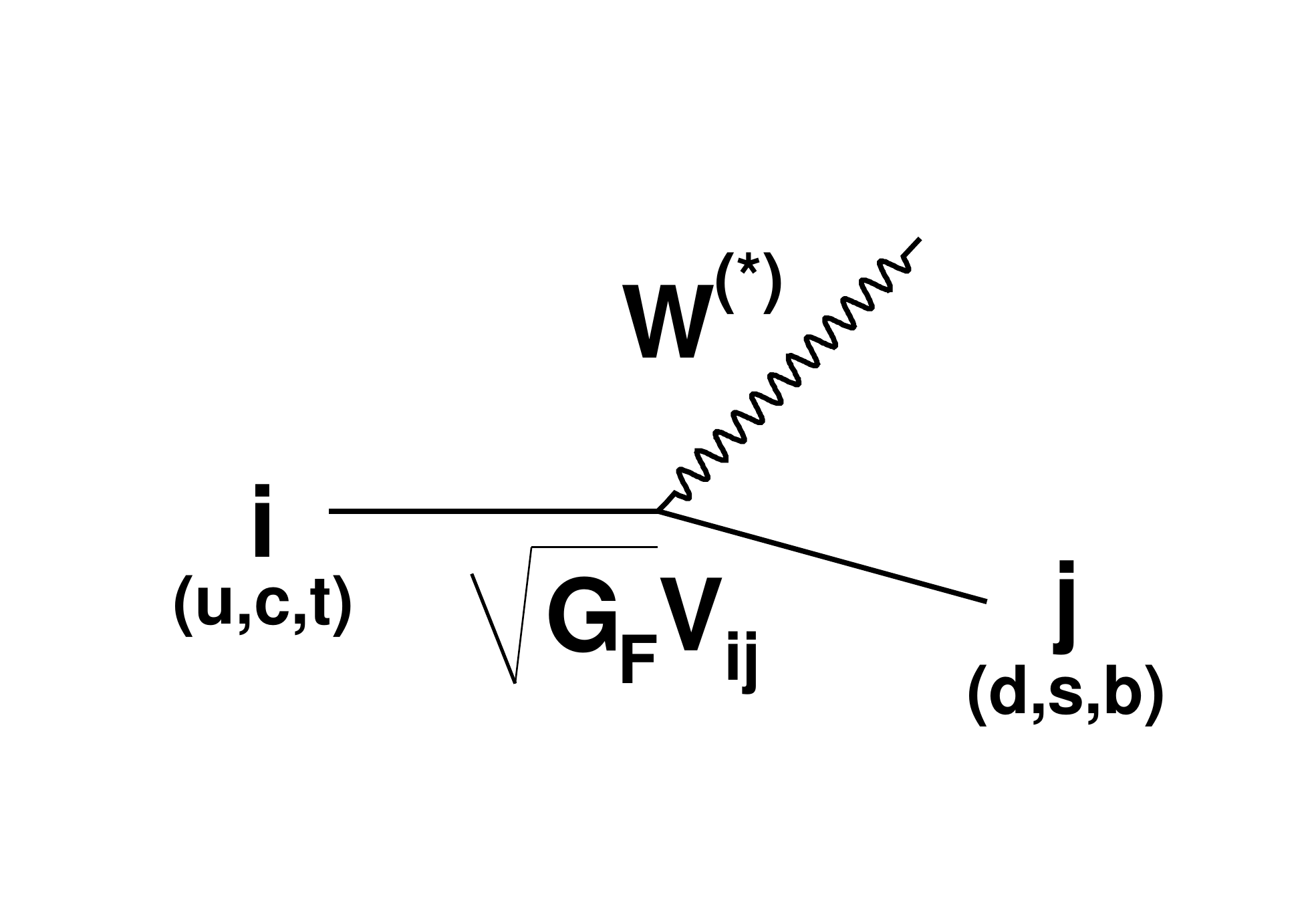}
    \includegraphics[width=0.63\columnwidth]{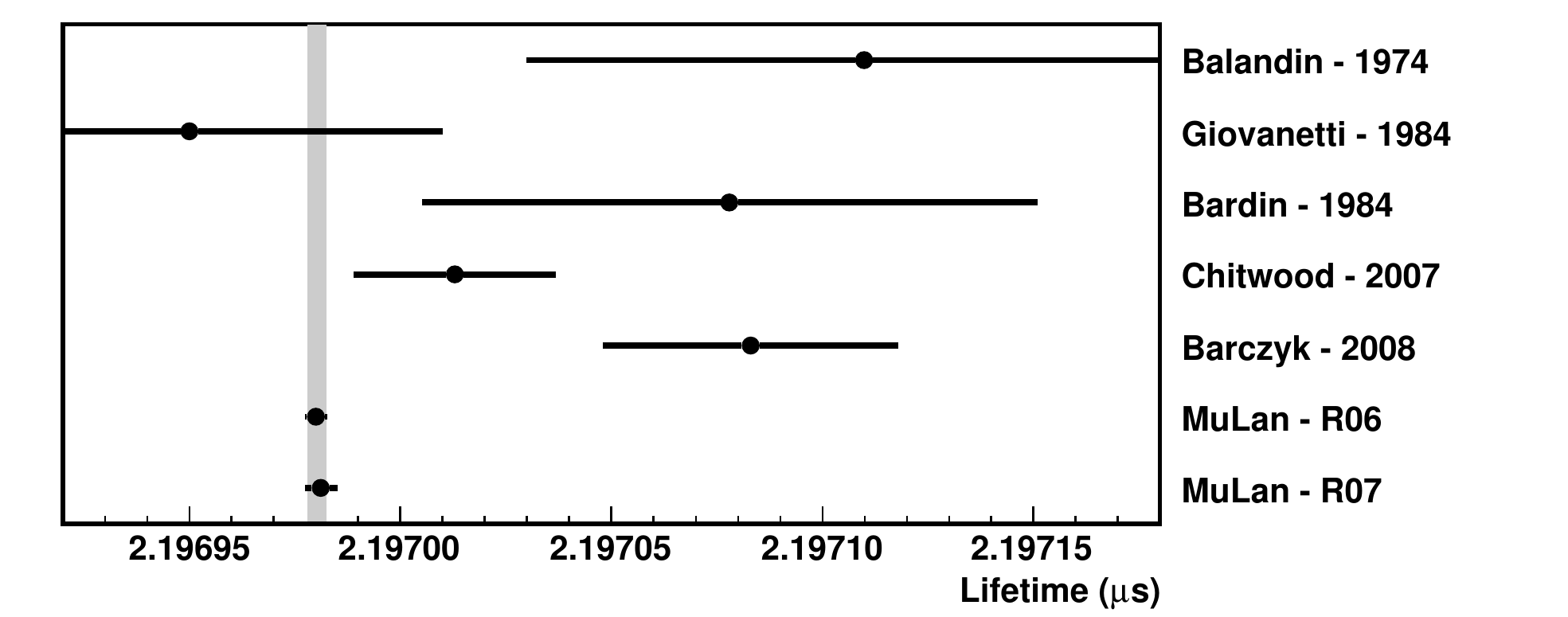}
    \caption{
      (Left) Diagram for a flavour-changing charged current interaction, the strength of which involves the Fermi constant $G_F$ and the CKM matrix element $V_{ij}$.
      (Right) Progress in the determination of the muon lifetime over the past forty years, from Ref.~\cite{Webber:2010zf}.
    }
    \label{fig:Vij}
  \end{center}
\end{figure} 

\section{\boldmath $CP$ conserving parameters -- magnitudes of CKM matrix elements}

\subsection{The Fermi constant}

As shown in Fig.~\ref{fig:Vij}, any absolute determination of the magnitude of a CKM matrix element requires knowledge of the Fermi constant, $G_F$.
The most precise information on $G_F$ is obtained from measurement of the lifetime of the muon, $\tau_\mu$, since
\begin{equation}
  \frac{1}{\tau_\mu} = \frac{G_F^2 m_\mu^5}{192\pi^3}\left(1+\Delta q\right) \, ,
\end{equation}
where $m_\mu$ is the mass of the muon (known to better than 50 parts per billion~\cite{Mohr:2008fa}) and $\Delta q$ accounts for phase-space, QED and hadronic radiative corrections (known to better than 1 part per million~\cite{Marciano:1999ih,Pak:2008qt}).
A recent measurement of the positive muon lifetime by the MuLan collaboration~\cite{Webber:2010zf}, set in its historical context in Fig.~\ref{fig:Vij}, gives
\begin{equation}
  \tau_{\mu^+} = \left( 2196980.3 \pm 2.2\right) {\rm ps} \, ,
\end{equation}
from which a determination of the Fermi constant to better than 1 part per million is obtained
\begin{equation}
  G_F = \left( 1.1663788 \pm 7 \right) \times 10^{-5} \, {\rm GeV}^{-2} \, .
\end{equation}

\subsection{Determination of $\left| V_{ud} \right|$}

\begin{figure}[ht]
  \begin{center}
    \includegraphics[width=0.38\columnwidth]{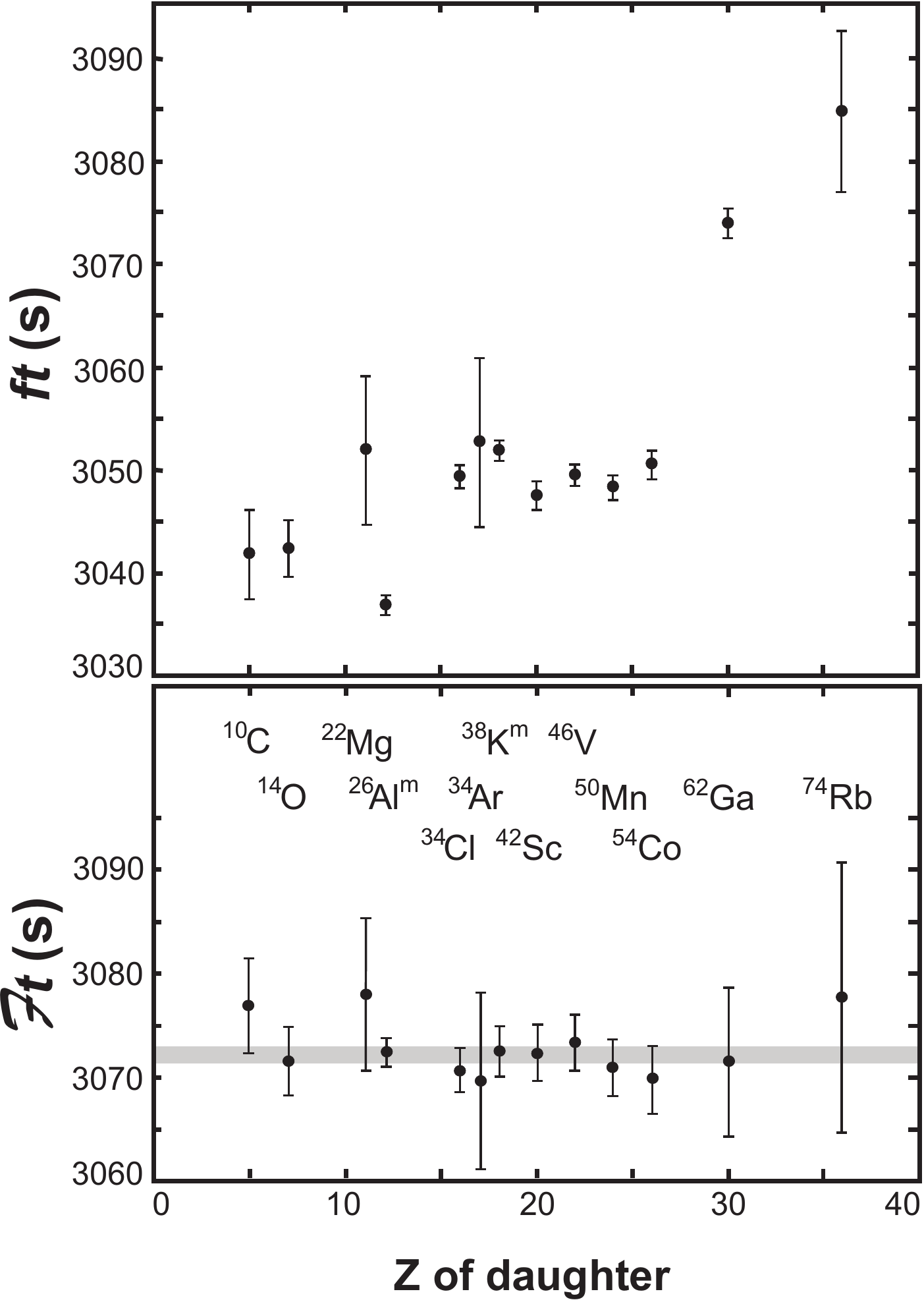}
    \caption{
      Uncorrected ($ft$) and corrected (${\cal F}t$) values obtained from different $0^+ \to 0^+$ transitions, from Ref.~\cite{Hardy:2008gy}.
    }
    \label{fig:Vud}
  \end{center}
\end{figure}

The most precise determination of $\left| V_{ud} \right|$ is from super-allowed $0^+ \to 0^+$ nuclear beta decays.
The relation 
\begin{equation}
  ft = \frac{K}{2G_F^2\left| V_{ud} \right|^2} \, ,
\end{equation}
where the parameters $f$ and $t$ are obtained from measurements of the energy gap and from the half-life and branching fraction, respectively, 
is expected to be nucleus-independent to first order ($K$ is a known constant, $K/(\hbar c)^6 = 2\pi^3 \hbar \ln 2/(m_e c^2)^5$).
However, as shown in Fig.~\ref{fig:Vud}, the precision is such that second-order effects related to the nuclear medium (radiative and isospin-breaking corrections) need to be accounted for.
This is achieved by obtaining a corrected quantity, labelled ${\cal F}t$~\cite{Hardy:2008gy}, that is confirmed to be constant to $3 \times 10^{-4}$ (see Fig.~\ref{fig:Vud}).
This allows $\left| V_{ud} \right|$ to be extracted,
\begin{equation}
  \left| V_{ud} \right| = 0.97425 \pm 0.00022 \, .
\end{equation}

Various alternative approaches allow determinations of $\left| V_{ud} \right|$, with different merits.
Nuclear mirror decays (transitions between states with nuclear isospin 1/2) and neutron lifetime measurements are sensitive to both vector and axial-vector couplings, and the latter does not require nucleus dependent or isospin breaking corrections to be known.
The current experimental status of the neutron lifetime is controversial~\cite{Nakamura:2010zzi,Pichlmaier:2010zz} (see also Refs.~\cite{0954-3899-36-10-104001,Paul:2009md}), but future experiments should reduce its uncertainty.
Determinations from pion beta decay (which has only vector couplings) have the smallest theoretical uncertainty, though significantly increased data samples would be necessary to approach the current sensitivity on $\left| V_{ud} \right|$.

\subsection{Determination of $\left| V_{us} \right|$}

\begin{figure}[ht]
  \begin{center}
    \includegraphics[width=0.30\columnwidth]{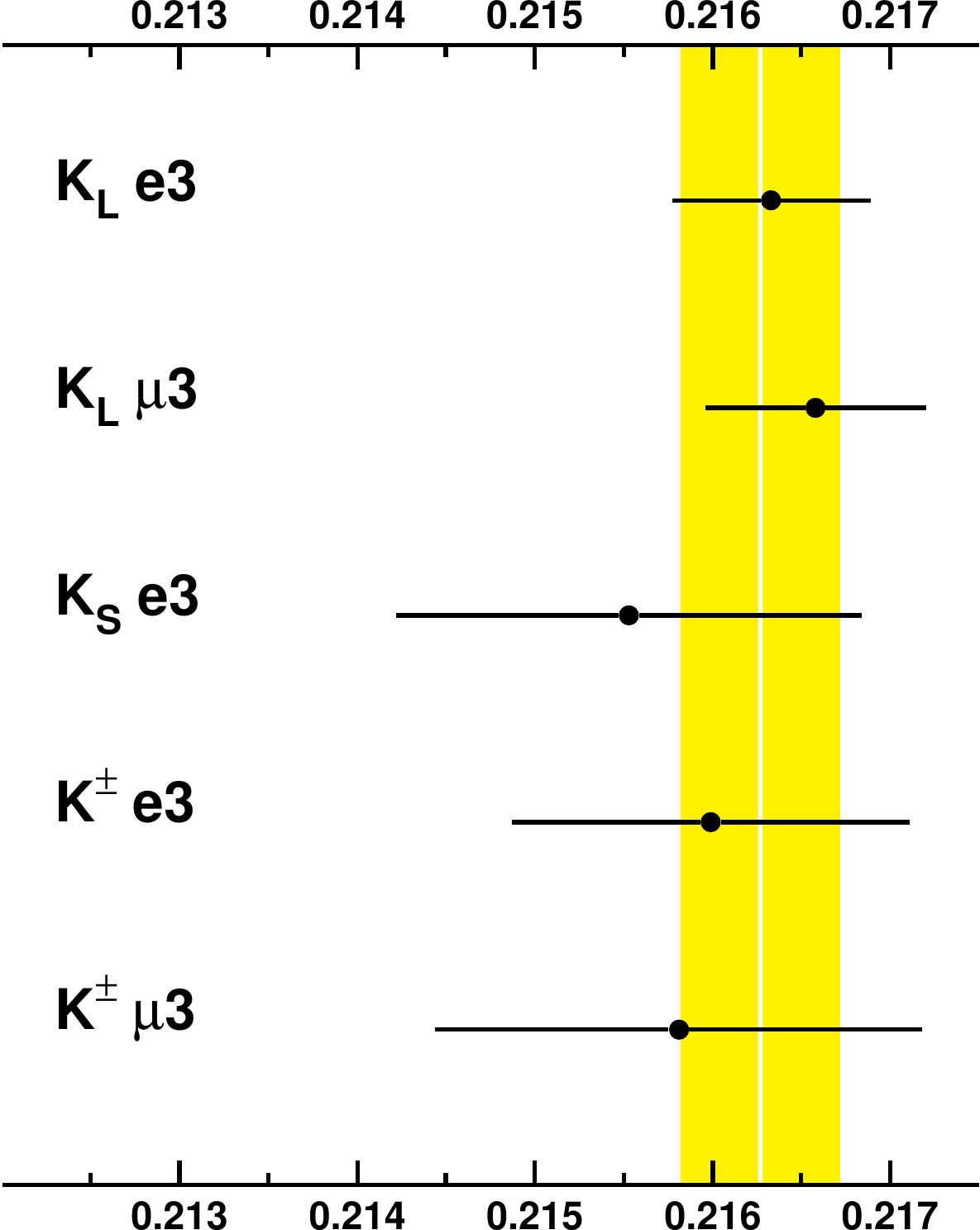}
    \hspace{8mm}
    \includegraphics[width=0.49\columnwidth]{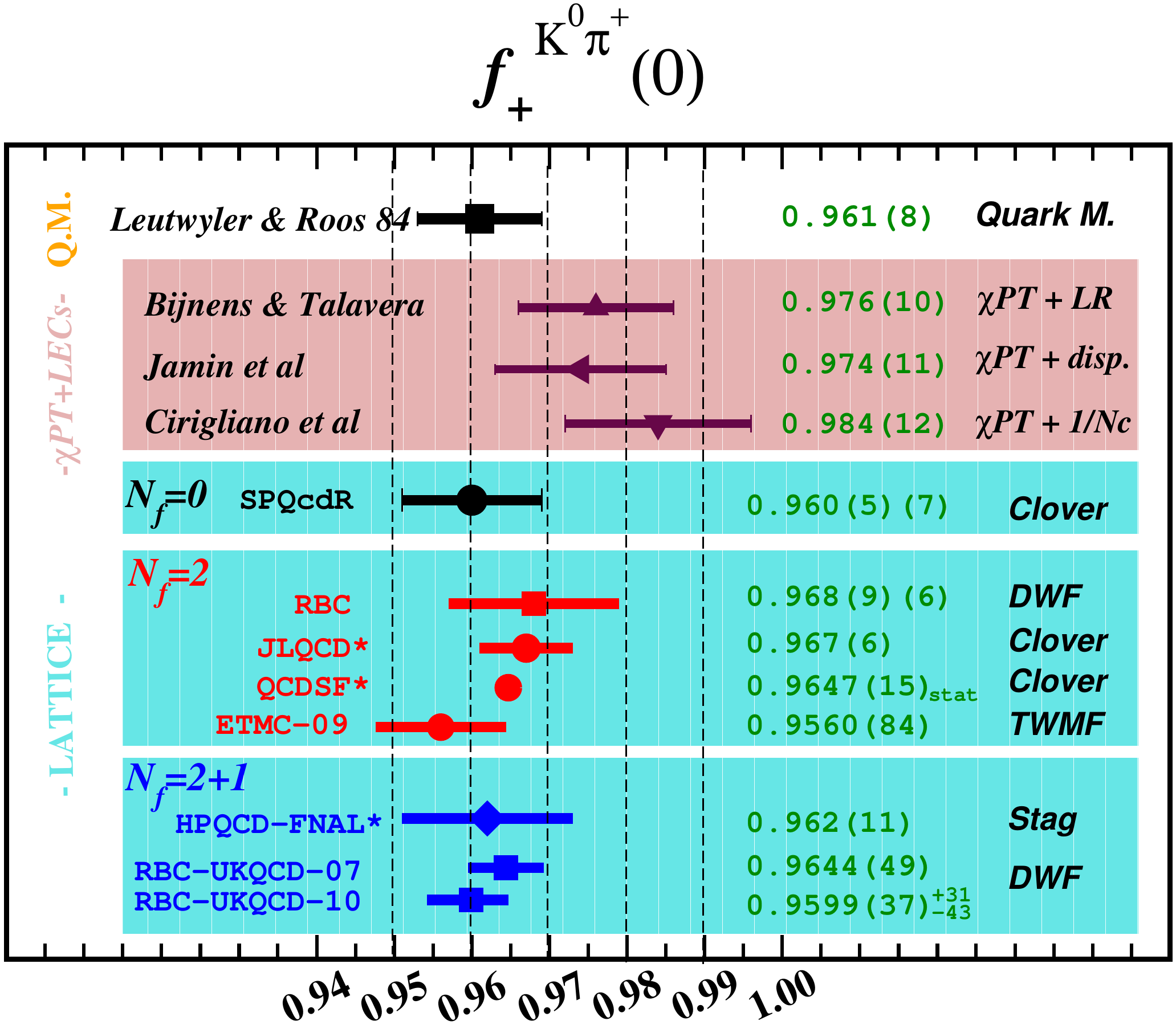}
    \caption{
      (Left) Values of $f_+(0) \left| V_{us} \right|$ obtained from different semileptonic kaon decays, giving an average $f_+(0) \left| V_{us} \right| = 0.2163 \pm 0.0005$.
      (Right) Calculations of $f_+^{K^0\pi^+}(0)$.
      From Ref.~\cite{Antonelli:2010yf}.
    }
    \label{fig:Vus}
  \end{center}
\end{figure}

The past few years have seen significant progress in the determination of $\left| V_{us} \right|$ from semileptonic kaon decays, as reviewed in Ref.~\cite{Antonelli:2010yf,Cirigliano:2011ny}.
Fig.~\ref{fig:Vus} summarises the values of $f_+(0) \left| V_{us} \right|$ determined experimentally, using data from the BNL-E865, KLOE, KTeV, ISTRA+ and NA48 experiments (the latest preliminary results from the NA48 collaboration~\cite{Veltri:2011zk} are not included), as well as the calculations of $f_+(0)$, mainly using lattice QCD techniques.
Using $f_+(0) = 0.959 \pm 0.005$~\cite{Boyle:2010bh}, the average value is obtained,
\begin{equation}
  \left| V_{us} \right| = 0.2254 \pm 0.0013 \, .
\end{equation}

\begin{figure}[ht]
  \begin{center}
    \includegraphics[width=0.45\columnwidth]{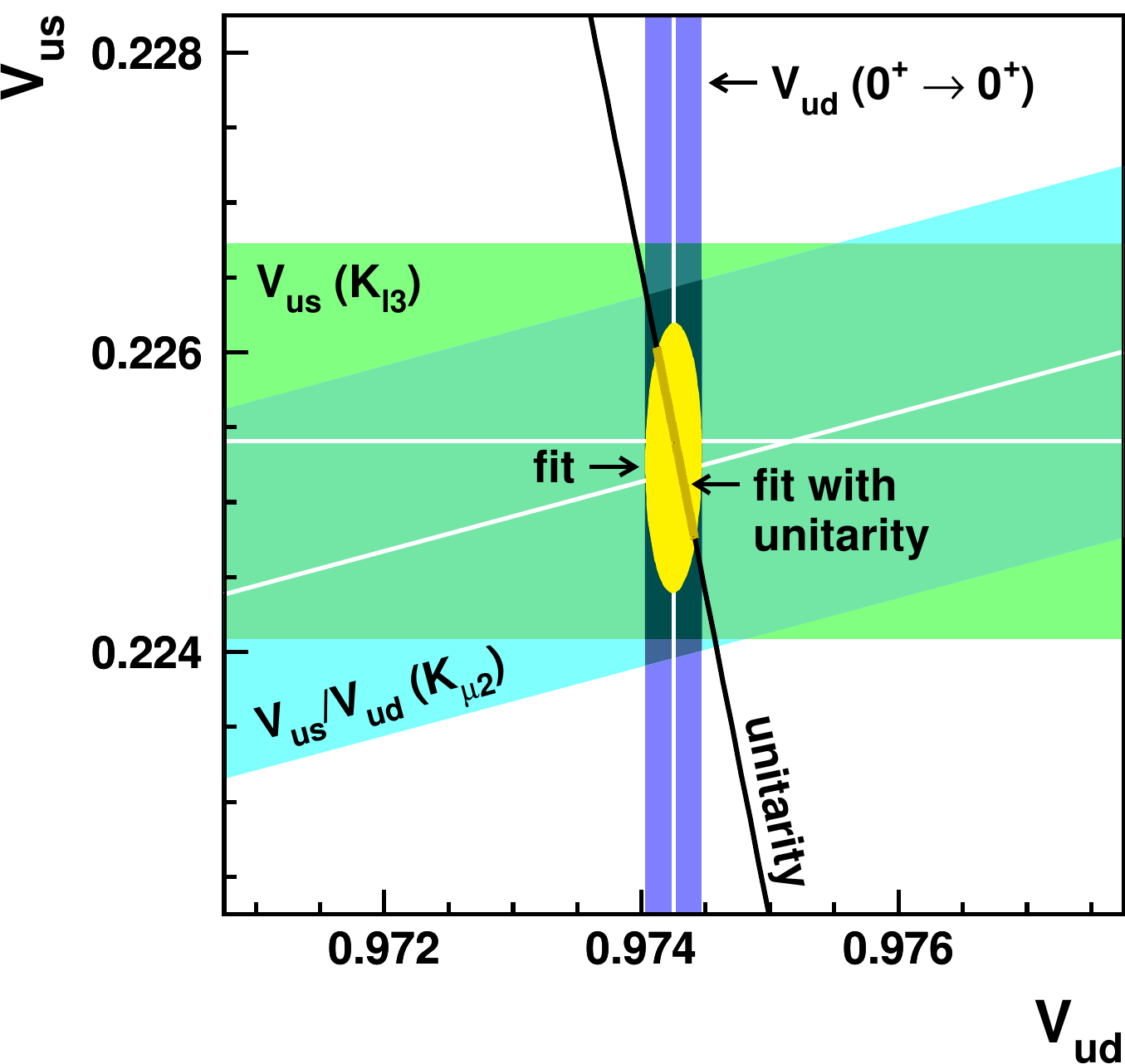}
    \caption{
      Combination of constraints on the magnitudes of the elements of the first row of the CKM matrix.
      From Ref.~\cite{Antonelli:2010yf}.
    }
    \label{fig:row1}
  \end{center}
\end{figure}

A result with comparable precision on the ratio of CKM matrix elements is obtained from the widths of leptonic kaon and pion decays.
The experimental data together with lattice QCD input, $f_K /f_\pi = 1.193 \pm 0.006$~\cite{Antonelli:2010yf}, and accounting for isospin violation~\cite{Cirigliano:2011tm} gives
\begin{equation}
  \left| V_{us} / V_{ud} \right| = 0.2316 \pm 0.0012 \, ,
\end{equation}
where both experimental and theoretical uncertainties are essentially uncorrelated with those in the average for $ \left| V_{us} \right|$ given above.
This then allows a comparison of the different determinations, as well as a test of the unitarity of the first row of the CKM matrix.
As shown in Fig.~\ref{fig:row1}, unitarity is found to hold to better than one part in $10^3$.
An alternative way of viewing this result is that the Fermi constant measured in the quark sector is consistent with the determination from the muon lifetime.
This is thus a beautiful demonstration of the universality of the weak interaction.

Alternative approaches to measure $\left| V_{us} \right|$ are possible using hyperon decays or hadronic tau lepton decays.
For the latter, the method relies on comparison of the inclusive strange and non-strange branching fractions.
These are determined experimentally from sums of exclusive measurements, and since not all decays have yet been measured, rely somewhat on extrapolations (see Ref.~\cite{Davier:2005xq} for a detailed review).
A recent study~\cite{Pich:2011cj} estimates the value from hadronic tau decays to be 
\begin{equation}
  \left| V_{us} \right| = 0.2166 \pm 0.0019 \,({\rm exp.}) \pm 0.0005 \,({\rm th.}) \, ,
\end{equation}
which is discrepant from the value from semileptonic kaon decays at the level of $3.7\,\sigma$.
Two important points are to be noted: firstly, the intrinsic theoretical uncertainty in this approach is very small; secondly, the central value may change as the $B$ factories complete their programmes of study of multibody hadronic tau decays.\footnote{
  As pointed out by A.~Hoecker at Lepton Photon, there is a significant discrepancy between the BaBar~\cite{Aubert:2007mh} and Belle~\cite{:2010tc} measurements of $\tau \to 3 \,{\rm tracks}\,+\, \nu$ branching fractions that should also be resolved.
}

\subsection{Determination of $\left| V_{cd} \right|$ and $\left| V_{cs} \right|$}

\begin{figure}[ht]
  \begin{center}
    \includegraphics[width=0.50\columnwidth]{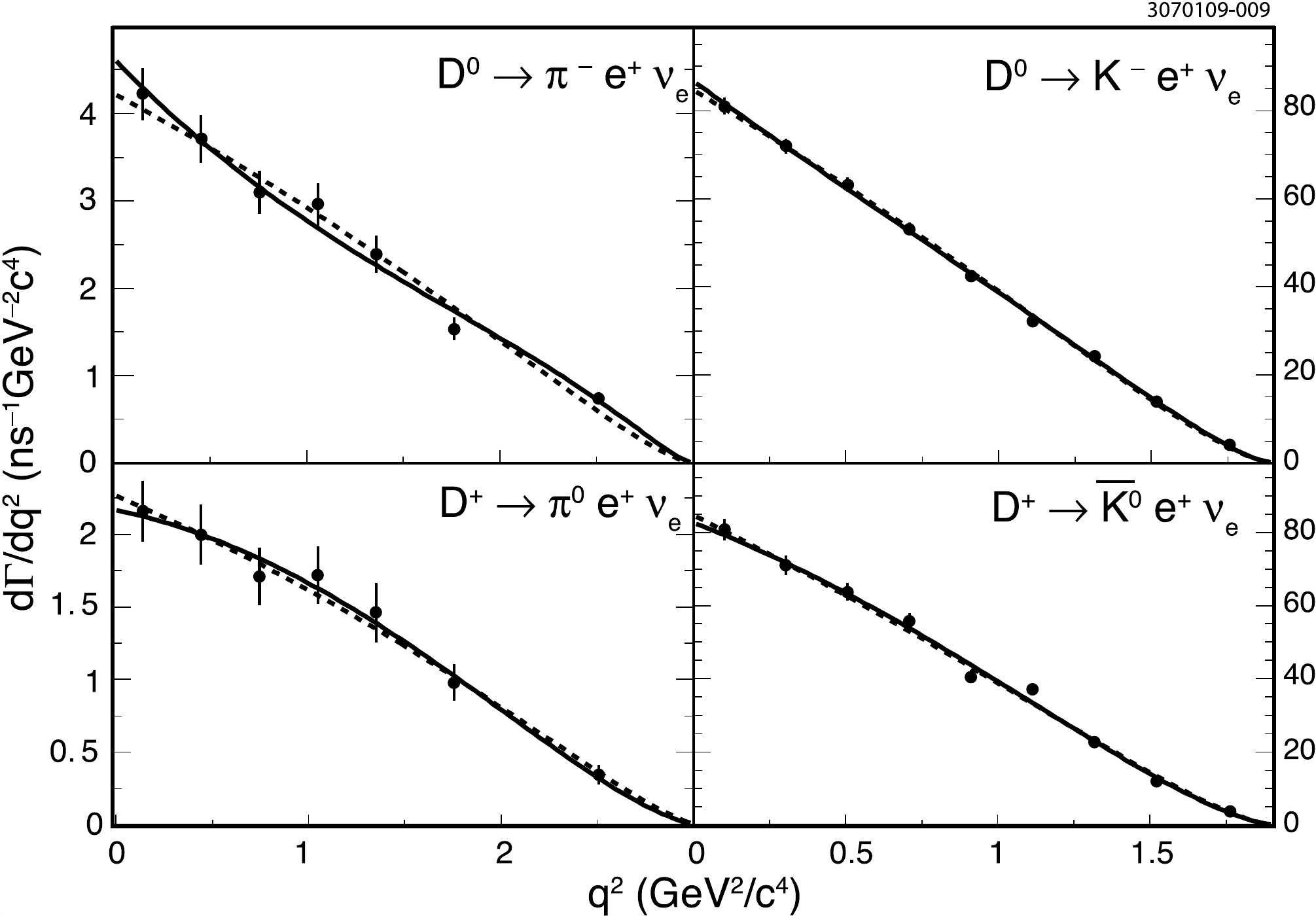}
    \caption{
      Differential branching fraction for semileptonic charm decays as a function of $e\nu$ invariant mass squared $q^2$, from CLEO-c~\cite{Besson:2009uv}.
      The results of fits to parametrised form factors are also shown.
    }
    \label{fig:cleoc-SL}
  \end{center}
\end{figure}

For several years, the benchmark determination of $\left| V_{cd} \right|$ has been that based on charm production in neutrino interactions, 
\begin{equation}
  \left| V_{cd} \right| = 0.230 \pm 0.011 \, .
\end{equation}
However, improved measurements of charm semileptonic decays, $D \to \pi l \nu$, from CLEO-c~\cite{Besson:2009uv}, provide the potential for further improvements.
The CLEO-c data is shown in Fig.~\ref{fig:cleoc-SL}.
A recent review~\cite{Laiho:2011uj} gives a value based on this approach
\begin{equation}
  \left| V_{cd} \right| = 0.234 \pm 0.007 \pm 0.002 \pm 0.025 \, ,
\end{equation}
where the last uncertainty is from lattice QCD determinations of the form factors~\cite{Na:2010uf,Aubin:2004ej}.
With reduced uncertainties from the lattice calculations, this promises to provide a more precise value of this CKM matrix element.\footnote{
  While these proceedings were in preparation, improved lattice calculations became available~\cite{Na:2011mc}.
}

Semileptonic charm decays, this time $D \to K l \nu$, also provide the most precise determination of $\left| V_{cs} \right|$.
Using inputs from CLEO-c~\cite{Besson:2009uv} (Fig.~\ref{fig:cleoc-SL}) and lattice QCD calculations~\cite{Na:2010uf}, the current value is
\begin{equation}
  \left| V_{cs} \right| = 0.961 \pm 0.011 \pm 0.024 \, ,
\end{equation}
where the uncertainties are experimental and from the lattice, respectively.

Leptonic charm meson decays provide an alternative approach to determine the magnitudes of these CKM matrix elements.
Their decay rates involve also the decay constants, which can be determined from lattice QCD, and helicity suppression factors, for example
\begin{equation}
  \Gamma \left( D_s^+ \to l^+ \nu \right) = \frac{G_F^2}{8\pi}f_{D_s^+}^2 m_l^2 M_{D_s^+} \left( 1 - \frac{m_l^2}{M_{D_s^+}^2} \right)^2 \left| V_{cs} \right|^2 \, .
\end{equation}
Significant improvements in the measurements of $D_s^+$ decays have come from BaBar~\cite{delAmoSanchez:2010jg}, Belle~\cite{:2007ws} and CLEO-c~\cite{Naik:2009tk}.
These are usually expressed in terms of $f_{D_s^+}$, using the value of $\left| V_{cs} \right|$ given above, and can be compared to the lattice QCD calculations.
Equally, this can be recast using the input from the lattice~\cite{Laiho:2009eu} to obtain 
\begin{equation}
  \left| V_{cs} \right| = 1.005 \pm 0.026 \pm 0.016 \, ,
\end{equation}
where the uncertainties are experimental and from the lattice, respectively.
It should be noted that a discrepancy that was apparent a few years ago (see, for example, Ref.~\cite{Dobrescu:2008er}) has disappeared.
Moreover, the dominant uncertainty is experimental, so improved measurements from BES and current or future $e^+e^-$ $B$ factory experiments would be welcome.

\subsection{Determination of $\left| V_{cb} \right|$ and $\left| V_{ub} \right|$}

Both exclusive and inclusive studies of semileptonic $B$ decays have been used to obtain $\left| V_{cb} \right|$ and $\left| V_{ub} \right|$
(for a detailed recent review, see Ref.~\cite{doi:10.1146/annurev.nucl.012809.104421}).
For the former, the review in the 2010 edition of the Particle Data Group review of particle physics~\cite{Nakamura:2010zzi} quotes a $2\,\sigma$ tension between the two determinations,
\begin{equation}
  \left| V_{cb} \right| \, ({\rm excl.}) = (38.7 \pm 1.1)\times 10^{-3} \, ,
  \hspace{3mm}
  \left| V_{cb} \right| \, ({\rm incl.}) = (41.5 \pm 0.7)\times 10^{-3} \, .
\end{equation}
Updated data from Belle on $B_d^0 \to D^{*-}l\nu$ decays~\cite{Dungel:2010uk} and improved lattice QCD calculations of the form-factor at zero recoil~\cite{Bailey:2010gb} reduce slightly both the uncertainty of the exclusive determination and the tension with the inclusive determination.

\begin{figure}[ht]
  \begin{center}
    \includegraphics[width=0.45\columnwidth,bb=0 320 285 640,clip=true]{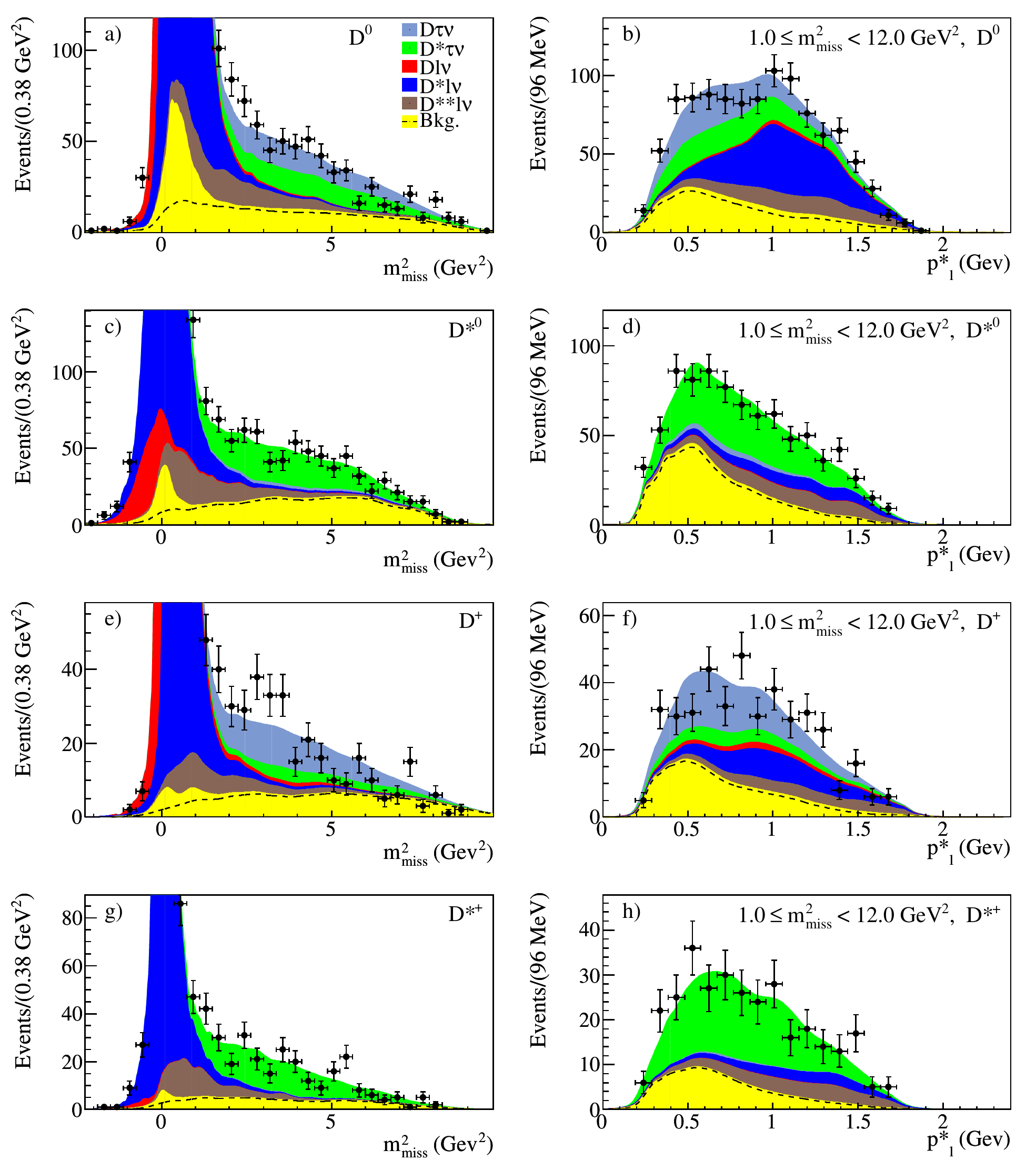}
    \includegraphics[width=0.45\columnwidth,bb=0 0 285 320,clip=true]{Fit_SigM2PlTail.pdf}
    \caption{
      Signal for $B \to D^{(*)}\tau\nu$ decays from BaBar~\cite{BaBarDtaunu}.
      Note that the large peaks are due to backgrounds from $D^{(*)}l\nu$ ($l=e,\mu$) decays, while the signals appears as tails to large values of the missing mass squared variable $m_{\rm miss}^2$.
    }
    \label{fig:BaBar-Dtaunu}
  \end{center}
\end{figure} 

It is also worth noting that the semitauonic decays $B \to D^{(*)}\tau\nu$ have recently been seen for the first time by BaBar~\cite{Aubert:2007dsa,:2009xy,BaBarDtaunu} (see Fig.~\ref{fig:BaBar-Dtaunu}) and Belle~\cite{Matyja:2007kt,Bozek:2010xy}. 
The rates of these decays depend on $\left| V_{cb} \right|$, but it is more common to measure their ratios relative to those for $B \to D^{(*)}l\nu$ ($l=e,\mu$) decays.
These ratios are precisely predicted in the SM, and are sensitive to potential contributions beyond the SM, for example from charged Higgs bosons.
The isospin averaged ratios are determined to be 
\begin{equation}
  \begin{array}{c@{\hspace{5mm}}c}
    R(D) = 0.456 \pm 0.053 \pm 0.056 \, , & 
    R^{\rm SM}(D) = 0.31 \pm 0.02 \, , \\
    R(D^*) = 0.325 \pm 0.023 \pm 0.027 \, , & 
    R^{\rm SM}(D^*) = 0.25 \pm 0.07 \, . \\
  \end{array}
\end{equation}
The excess over the SM is about $1.8\,\sigma$ (see also Ref.~\cite{kwon}, where determinations of $\left| V_{ub} \right|$ from leptonic $B$ decays are also discussed).

\begin{figure}[ht]
  \begin{center}
    \includegraphics[width=0.45\columnwidth]{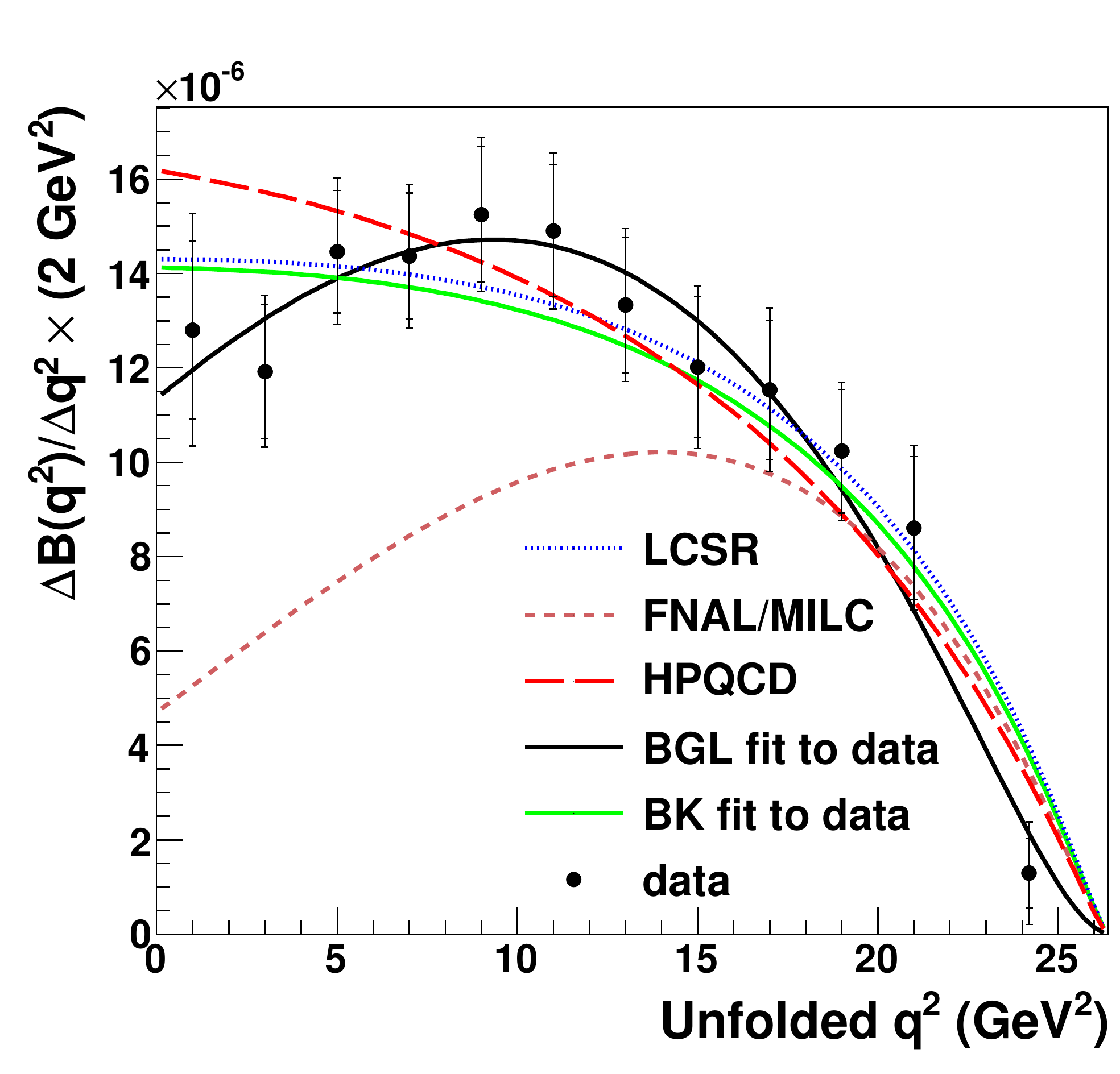}
    \includegraphics[width=0.45\columnwidth]{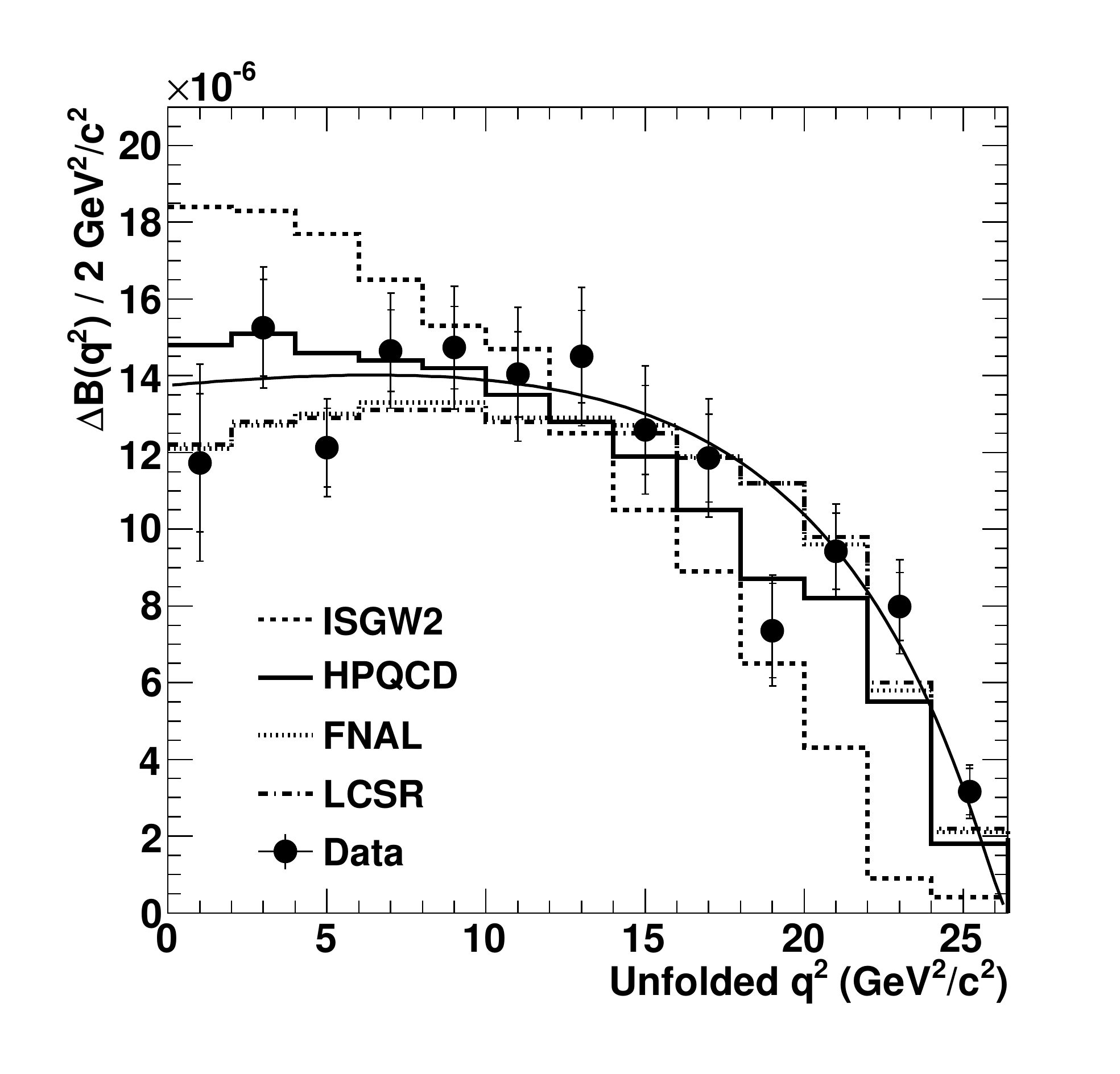}
    \caption{
      Differential branching fractions of $B_d^0 \to \pi^- l^+\nu$ decays as a function of $l^+\nu$ invariant mass squared $q^2$ from (left) BaBar~\cite{delAmoSanchez:2010zd} and (right) Belle~\cite{Ha:2010rf}.
    }
    \label{fig:Vub}
  \end{center}
\end{figure}

The $b \to u l \nu$ decays can similarly be used to obtain measurements of $\left| V_{ub} \right|$ by either exclusive or inclusive methods.
Most recent progress has been on the exclusive $B \to \pi l \nu$ decays, where new results have become available from both BaBar~\cite{delAmoSanchez:2010zd,:2010uj} and Belle~\cite{Ha:2010rf}, as shown in Fig.~\ref{fig:Vub}.
The updated HFAG~\cite{Asner:2010qj} average is~\cite{UrquijoEPS}
\begin{equation}
  \left| V_{ub} \right| = (3.26 \pm 0.30) \times 10^{-3} \, ,
\end{equation}
where the dominant source of uncertainty is from the lattice QCD calculations of the form factors~\cite{Bernard:2009ke}.

As was the case for $\left| V_{cb} \right|$, there is a tension between inclusive and exclusive determinations (the world average value using the inclusive approach is $\left| V_{ub} \right| = (4.27 \pm 0.38) \times 10^{-3}$.
Although some commentators have pointed out that the large amount of theoretical work dedicated to the extraction of $\left| V_{ub} \right|$ may have led to an underestimation of the uncertainties~\cite{Altarelli:2011vt}, it is this author's view that {\it more} theoretical attention is necessary to resolve the situation.
On the exclusive side, improvements in lattice QCD calculations can be expected, while on the inclusive side an initiative to reduce uncertainties using global fits is underway~\cite{Bernlochner:2011di}.

\section{\boldmath $CP$ violating parameters -- angles of the Unitarity Triangle and other phases}

As is widely known, $CP$ violation is one of the three ``Sakharov conditions''~\cite{Sakharov:1967dj} necessary for the evolution of a baryon asymmetry in the Universe.
Moreover, the SM $CP$ violation, encoded in the CKM matrix, is not sufficient to explain the observed asymmetry.
Therefore, there must be more sources of matter-antimatter asymmetry in nature.
These could arise in almost any conceivable extension of the SM, such as in an extended quark sector, in the lepton sector (leptogenesis), from anomalous gauge couplings, in an extended Higgs sector, and so on.
While all of these must be investigated, testing the consistency of the CKM mechanism in the quark sector provides the best chance to find new sources of $CP$ violation in the short term.

Although the understanding of $CP$ violation has advanced dramatically over the past decade, it is important to realise that it remains a rarely observed phenomenon.
To date, it is only been observed (with $>5\,\sigma$ significance) in the $K^0$ and $B_d^0$ systems.
(Discussions of searches for $CP$ violation in $D^0$ and $B_s^0$ mixing can be found in Refs.~\cite{raven,vanKooten}.)
In the $B$ system, the only $5\,\sigma$ significant measurements are of the parameters $\sin(2\beta)$ from $J/\psi K_{S,L}$ and similar decays, from BaBar~\cite{:2009yr} and Belle~\cite{Belle-sin2beta};
$S(\eta^\prime K_{S,L})$, from BaBar~\cite{:2008se} and Belle~\cite{Chen:2006nk};
$S(\pi^+\pi^-)$, from BaBar~\cite{Aubert:2007mj} and Belle~\cite{Ishino:2006if};
$C(\pi^+\pi^-)$, from Belle~\cite{Ishino:2006if};
and $A_{CP}(K^+\pi^-)$ from BaBar~\cite{Aubert:2007mj}, Belle~\cite{Belle-Kpi} and LHCb~\cite{LHCb-CONF-2011-042} (see also Ref.~\cite{kwon} on this last topic).
The LHCb result on $B_d^0 \to K^+\pi^-$ is thus the first $5\,\sigma$ observation of $CP$ violation in the $B$ system at a hadron collider experiment.

$CP$ violation results are often expressed in terms of the so-called Unitarity Triangle, which is a graphical representation of one of the relations implied by the unitarity of the CKM matrix,
\begin{equation}
  V_{ud}V^*_{ub} + V_{cd}V^*_{cb} + V_{td}V^*_{tb} = 0 \, .
\end{equation}
The angles of this triangle are usually denoted $(\alpha, \beta, \gamma)$, while its apex (after normalising so that its base is unit length along the real axis) is given in terms of the Wolfenstein parameters $(\bar{\rho}, \bar{\eta})$~\cite{Wolfenstein:1983yz,Buras:1994ec}.

\subsection{Searches for $CP$ violation in the charm sector}

Almost all $CP$ violation effects in the charm system are expected to be negligible in the SM.
This therefore provides an excellent testing ground to look for unexpected effects.
Various searches for direct $CP$ violation effects (studies of mixing and indirect $CP$ violation are discussed in Ref.~\cite{vanKooten}) have been carried out recently, for example in $D_{(s)}^+ \to K_S \pi^+$ and $K_S K^+$ decays~\cite{Ko:2010ng,delAmoSanchez:2011zza}, in triple product asymmetries in four-body hadronic decays~\cite{delAmoSanchez:2010xj,Lees:2011dx} and in Dalitz plot asymmetries in three-body decays~\cite{:2011cw}.
At the time of Lepton Photon, no significant signal for $CP$ violation in charm had yet been seen; although the world average asymmetry in $D^+ \to K_S \pi^+$ is more than $3\,\sigma$ from zero~\cite{Asner:2010qj} this is consistent with originating from the $CP$ violation in the neutral kaon system (see Ref.~\cite{Grossman:2011zk} and references therein).
However, while these proceedings were being prepared, LHCb announced a $3.5\,\sigma$ signal for the difference in time-integrated $CP$ asymmetries between $D^0 \to K^+K^-$ and $D^0 \to \pi^+\pi^-$ decays~\cite{:2011in} (CDF have also released less precise results on the same observable~\cite{Aaltonen:2011se}).

\subsection{Measurement of $\sin(2\beta)$}

\begin{figure}[ht]
  \begin{center}
    \includegraphics[width=0.45\columnwidth]{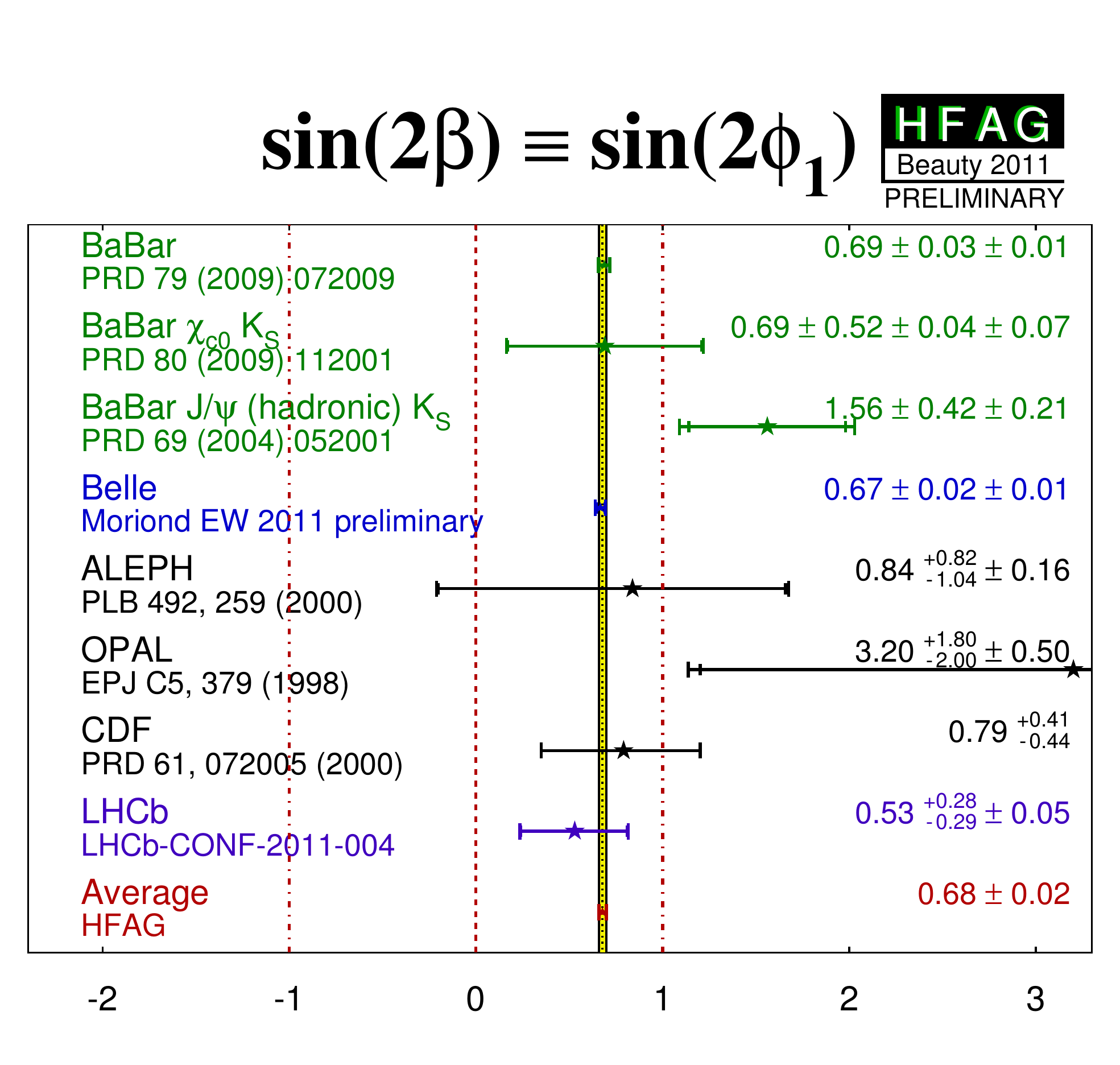}
    \includegraphics[width=0.42\columnwidth]{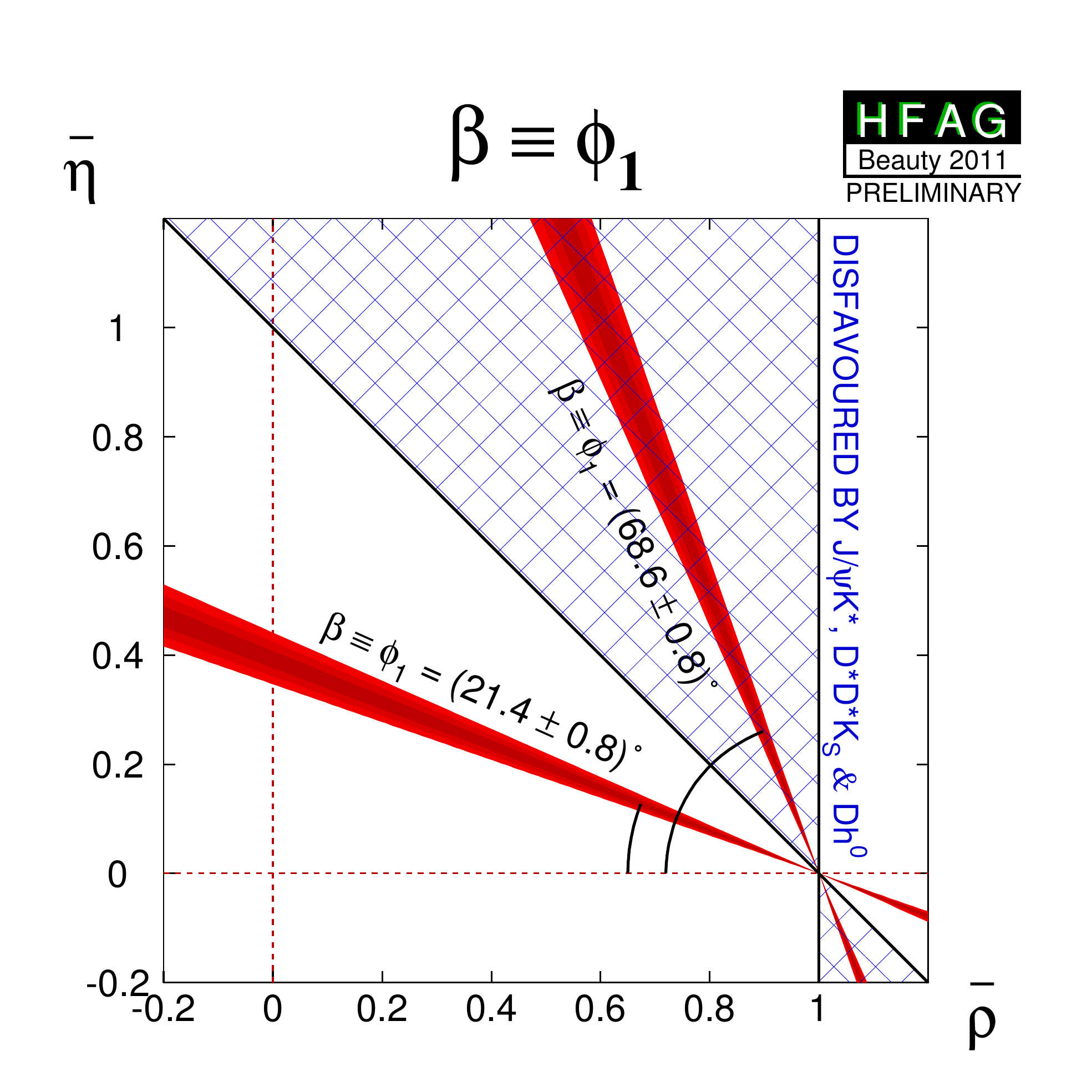}
    \caption{
      (Left) Compilation of results on $\sin(2\beta)$ from $B_d^0 \to J/\psi K_{S,L}$ ({\it etc.})~\cite{Asner:2010qj}.
      (Right) Corresponding constraint on $\bar{\rho}$--$\bar{\eta}$ plane.
    }
    \label{fig:sin2beta}
  \end{center}
\end{figure}

Both $e^+e^-$ $B$ factory experiments, BaBar and Belle, have completed data taking.  
The result on $\sin(2\beta)$ from $B_d^0 \to J/\psi K_{S,L}$ ({\it etc.}) with BaBar's final data set (445 million $B\bar{B}$ pairs) has been published~\cite{:2009yr}, while preliminary results following a reprocessing of the Belle data (772 million $B\bar{B}$ pairs) are available~\cite{Belle-sin2beta}.
A first analysis from LHCb is also available~\cite{LHCb-CONF-2011-004}.
The results are compiled in Fig.~\ref{fig:sin2beta}.
At the level of precision that the experiments are reaching, it is important to check for effects that may perturb the na\"ive SM expectation $S(J/\psi K_{S,L}) = -\eta_{CP} \sin(2\beta)$, where $\eta_{CP}$ is the $CP$ eigenvalue of the final state.
This can be done using channels that are related by flavour symmetries -- $B_d^0 \to J/\psi \pi^0$ (related by SU(3)) or $B_s^0 \to J/\psi K_S^0$ (related by U-spin).
First observations of the latter decay have recently been reported by CDF and LHCb~\cite{Aaltonen:2011sy,LHCb-CONF-2011-048}, suggesting that this approach will be possible with larger datasets.

\begin{figure}[ht]
  \begin{center}
    \includegraphics[width=0.45\columnwidth]{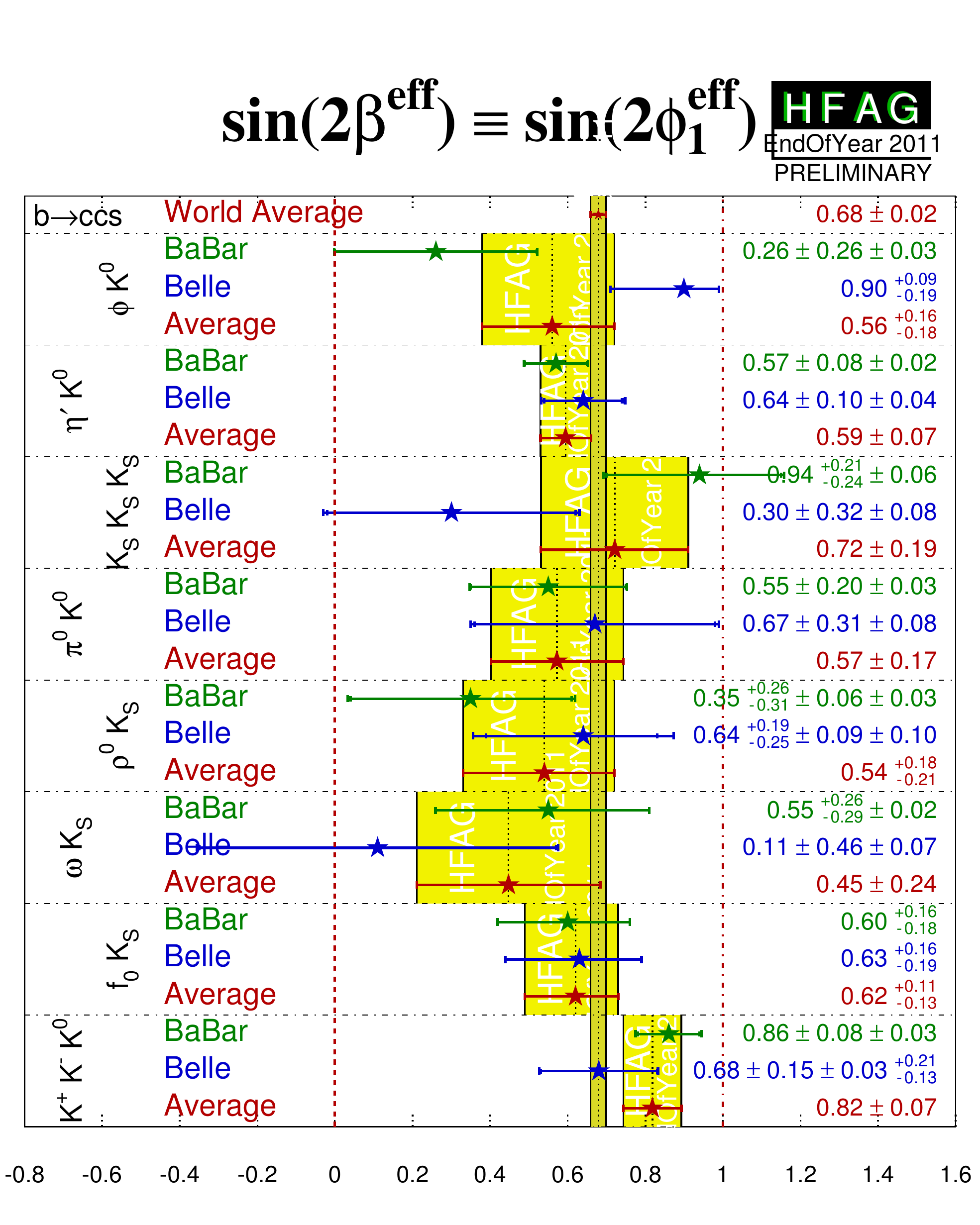}
    \includegraphics[width=0.45\columnwidth]{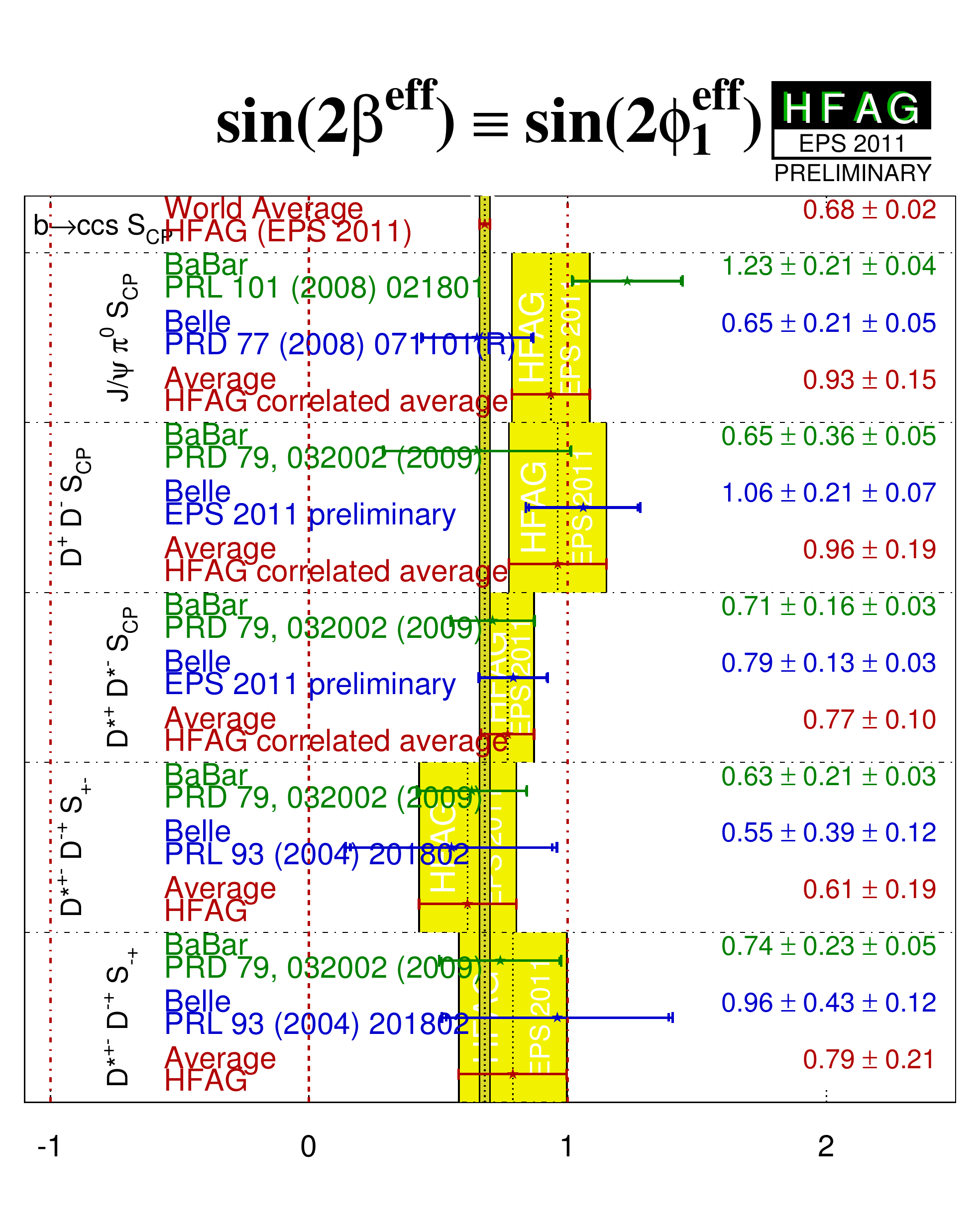}
    \caption{
      Compilation of results on $\sin(2\beta^{\rm eff})$ from (left) $b \to q\bar{q}s$ and (right) $b \to c\bar{c}d$ transitions~\cite{Asner:2010qj}.
    }
    \label{fig:qqsccd}
  \end{center}
\end{figure}

The $B$ factories have carried out a substantial programme of alternative measurements of $\sin(2\beta)$ using different quark level transitions, such as $b \to q\bar{q}s$ ($q=u,d,s$; {\it e.g.} $B_d^0 \to \eta^\prime K_S^0$) and $b \to c\bar{c}d$ ({\it e.g.} $B_d^0 \to D^+D^-$).
Compilations are shown in Fig.~\ref{fig:qqsccd}.
A few years ago, hints of deviations were apparent between the value of $\sin(2\beta^{\rm eff})$ measured in $b\to q\bar{q}s$ transitions and the reference value from $b \to c\bar{c}s$.
These have diminished with the latest data, but effects of non-SM contributions at the ${\cal O}(10\%)$ level are not ruled out.
One notable update is the new Belle result on $B_d^0 \to D^+D^-$~\cite{Belle-DD}, which improves the consistency between the results of the two $B$ factories as well as with the SM.

\subsection{Measurement of $\alpha$}

The unitarity triangle angle $\alpha$ is constrained by measurements of, and isospin relations between $B \to \pi\pi,\,\rho\pi$ and $\rho\rho$ decays~\cite{Gronau:1990ka,Snyder:1993mx}.
The situation has been stable for the last few years, though the final results from both $B$ factory experiments in all three systems are still awaited.
Combining all available information, the world average is~\cite{Charles:2004jd}
\begin{equation}
  \alpha = \left( 89.0\,^{+4.4}_{-4.2} \right)^\circ \, .
\end{equation}
Since the average is dominated by results from the $\rho\rho$ system two small comments are in order.
First, the apparently high branching fraction of $B^+ \to \rho^+\rho^0$, which comes essentially from a single measurement~\cite{Aubert:2009it} stretches the isospin triangle and reduces the uncertainty.
Secondly, analyses to date, while allowing $CP$ violation in the rates, have assumed the longitudinal polarisation fraction is the same for $B$ and $\bar{B}$ -- but the most general analysis would allow a difference between the two.

\subsection{Measurement of $\gamma$}

The angle $\gamma$ is unique among $CP$ violating observables, in that it can be determined using tree-level processes only, exploiting the interference between (typically) $b \to c\bar{u}d$ and $b \to u\bar{c}d$ transitions that occurs when the process involves a neutral $D$ meson reconstructed in a final state accessible to both $D^0$ and $\bar{D}^0$ decays.
It therefore provides a SM benchmark, and its precise measurement is crucial in order to disentangle any non-SM contributions to other processes, via global CKM fits.

\begin{figure}[ht]
  \begin{center}
    \includegraphics[width=0.60\columnwidth,bb=0 260 595 510,clip=true]{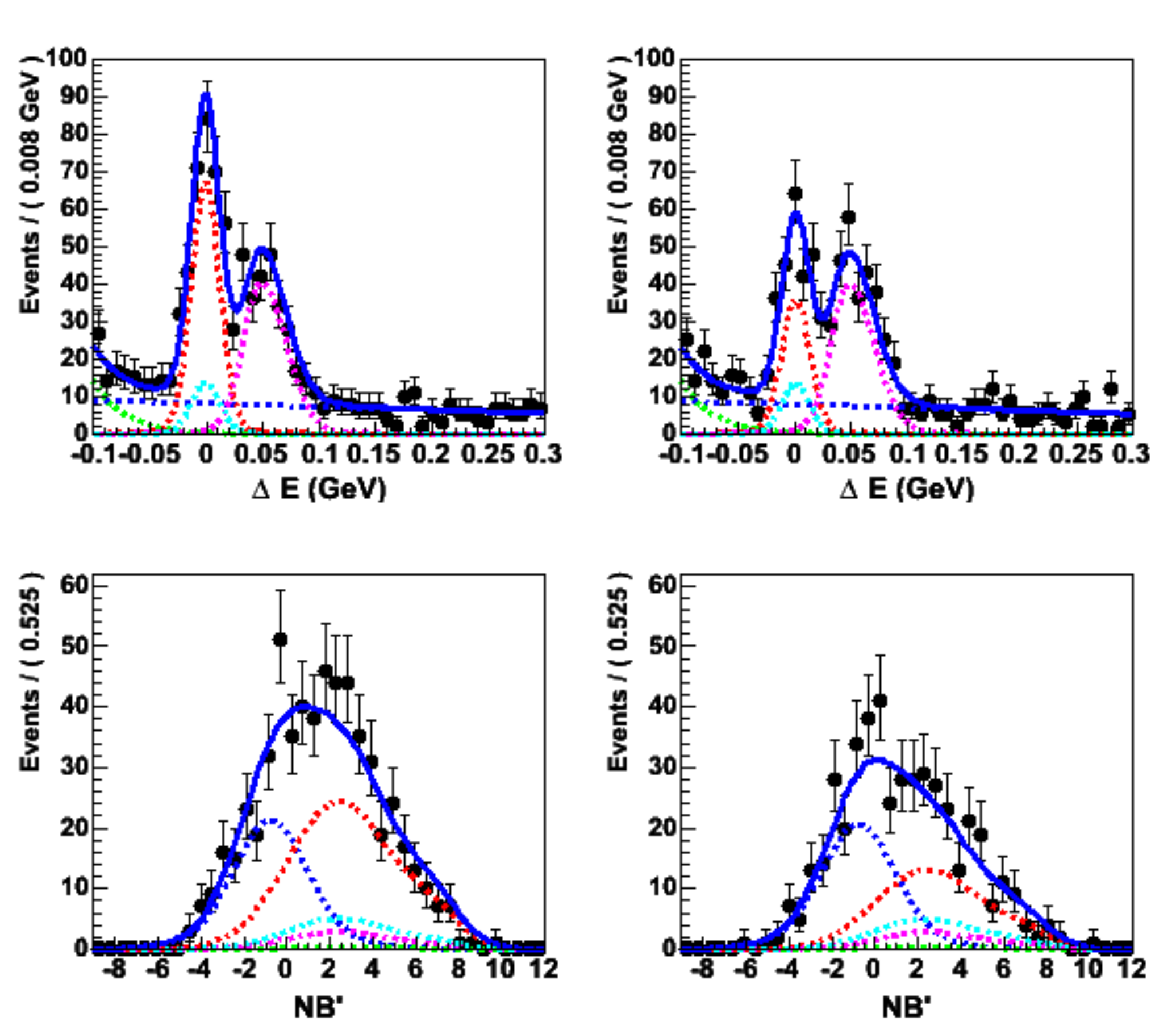}
    \includegraphics[width=0.60\columnwidth,bb=0 260 595 510,clip=true]{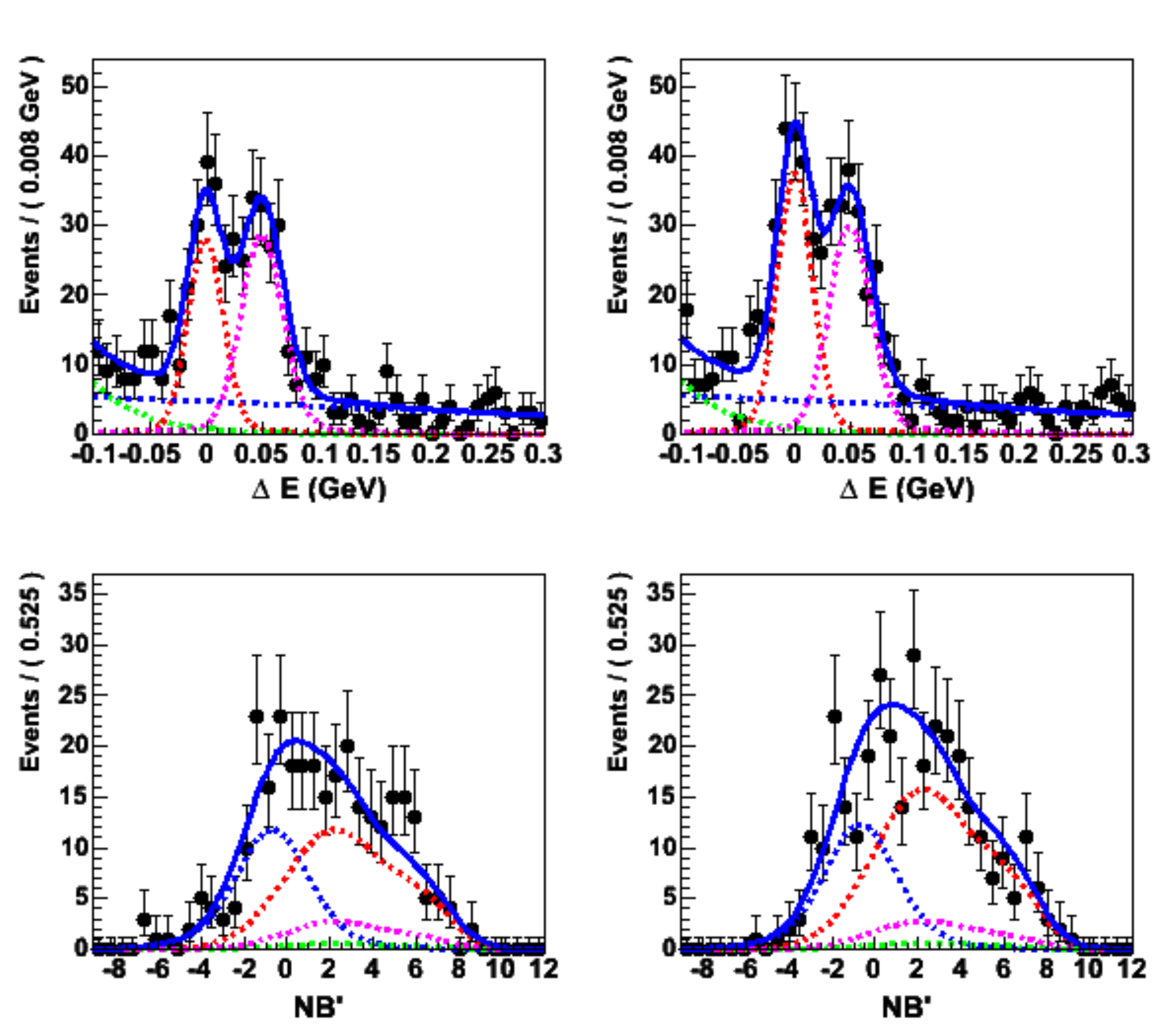}
    \caption{
      Signals for $B^\pm \to D_{CP}K^\pm$ decays from Belle~\cite{Belle-DKGLW}.
      (Top two plots) $D$ decays to $CP$-even final states ($K^+K^-$, $\pi^+\pi^-$).
      (Bottom two plots) $D$ decays to $CP$-odd final states ($K_S^0\pi^0$, $K_S^0\eta$).
      In each pair of plots the left (right) figure is for $B^-$ ($B^+$) decays.
      The plotted variable, $\Delta E$, peaks at zero for signal decays, while background from $B^\pm \to D\pi^\pm$ appears as a satellite peak at positive values.
    }
    \label{fig:BelleDKGLW}
  \end{center}
\end{figure}

\begin{figure}[ht]
  \begin{center}
    \includegraphics[width=0.45\columnwidth]{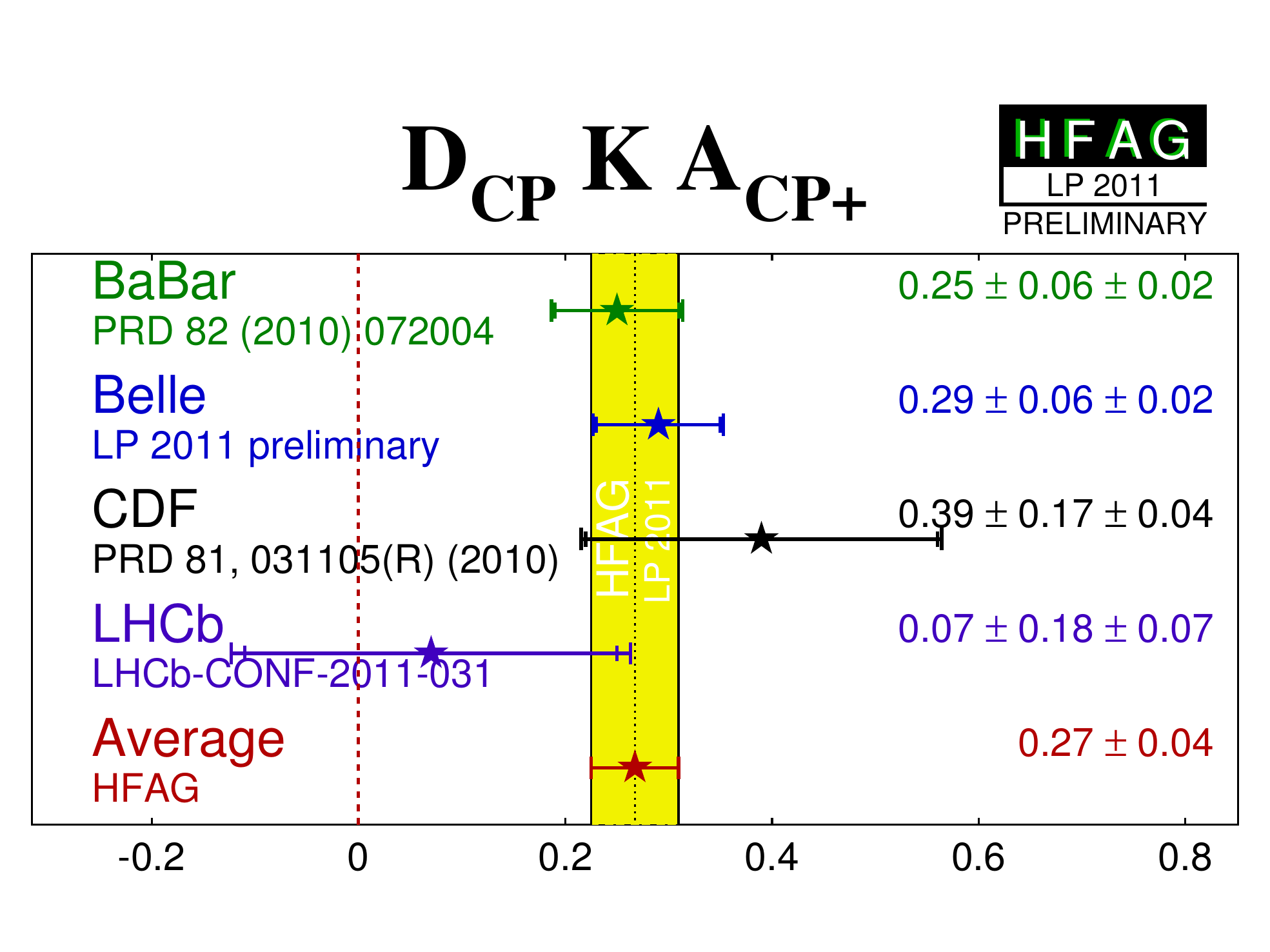}
    \includegraphics[width=0.45\columnwidth]{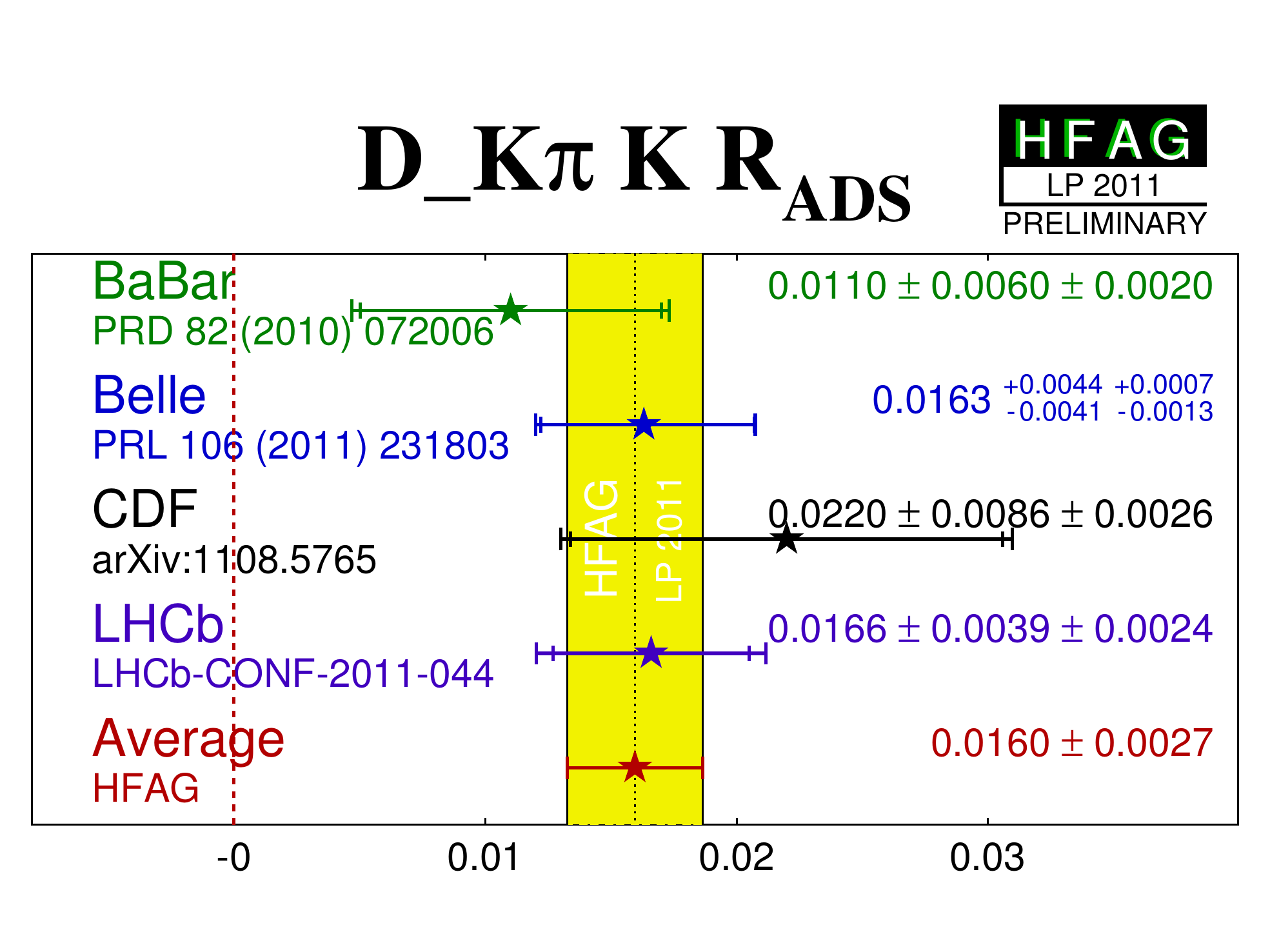}
    \caption{
      Compilations of results, with world averages, for $B^\pm \to DK^\pm$ decays in (left) GLW and (right) ADS modes~\cite{Asner:2010qj}.
    }
    \label{fig:HFAG-DK}
  \end{center}
\end{figure}

Several different $D$ decay final states have been studied in order to maximise the sensitivity to $\gamma$.
The archetype is the use of $D$ decays to $CP$ eigenstates, the so-called GLW method~\cite{Gronau:1990ra,Gronau:1991dp}.
New results with this approach have recently become available from BaBar~\cite{delAmoSanchez:2010ji}, CDF~\cite{Aaltonen:2009hz} and LHCb~\cite{LHCb-CONF-2011-031}, while the very latest results from Belle~\cite{Belle-DKGLW} are shown in Fig.~\ref{fig:BelleDKGLW}.
The world average for the $CP$ asymmetry in the processes involving $CP$-even $D$ decay final states, including all these new results and illustrated in Fig.~\ref{fig:HFAG-DK} (left), shows that $CP$ violation in $B^\pm \to DK^\pm$ decays is clearly established, though no single measurement exceeds $5\,\sigma$ significance.

\begin{figure}[ht]
  \begin{center}
    \includegraphics[width=0.75\columnwidth,bb=0 168 600 340,clip=true]{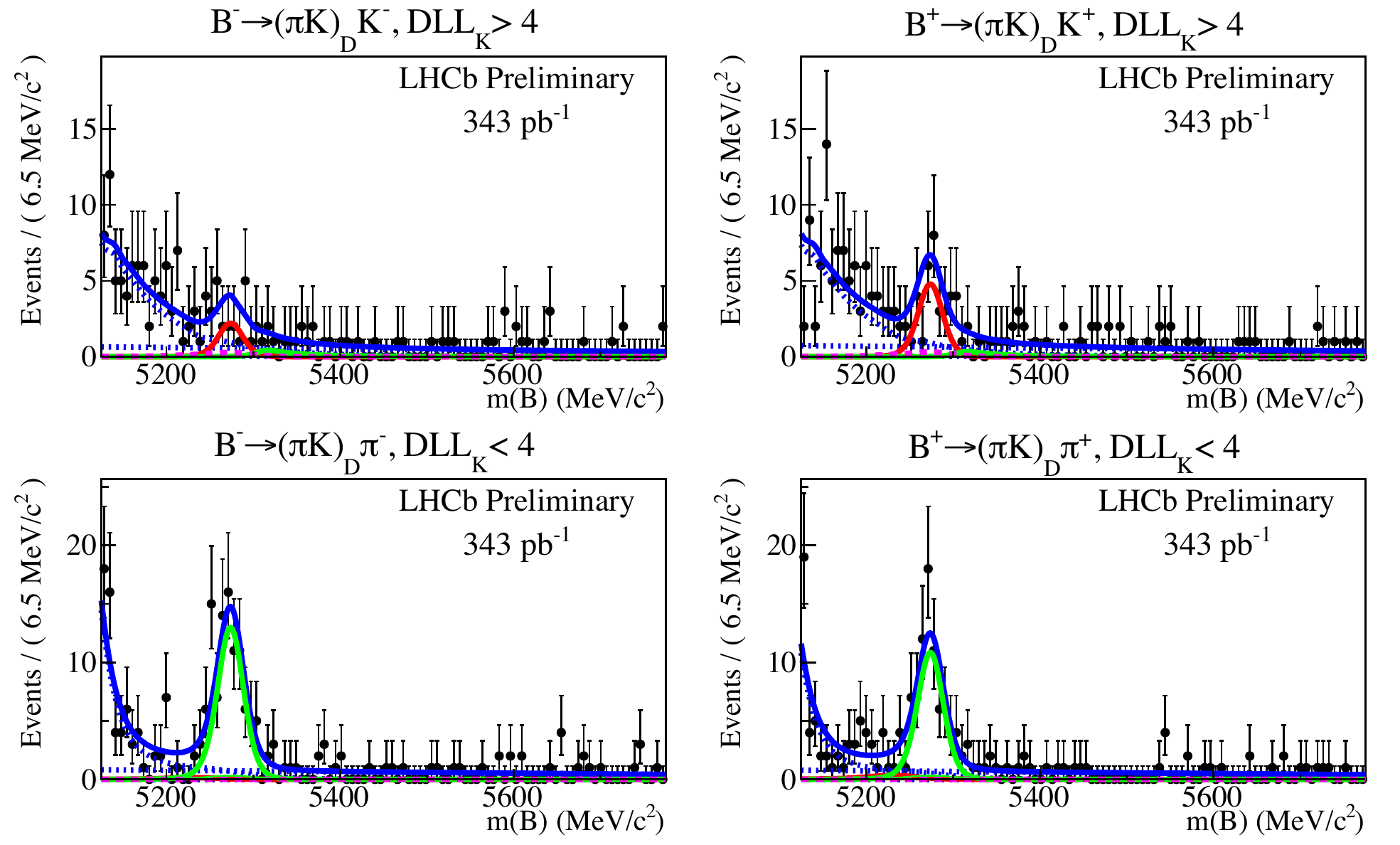}
    \includegraphics[width=0.75\columnwidth,bb=0 168 600 340,clip=true]{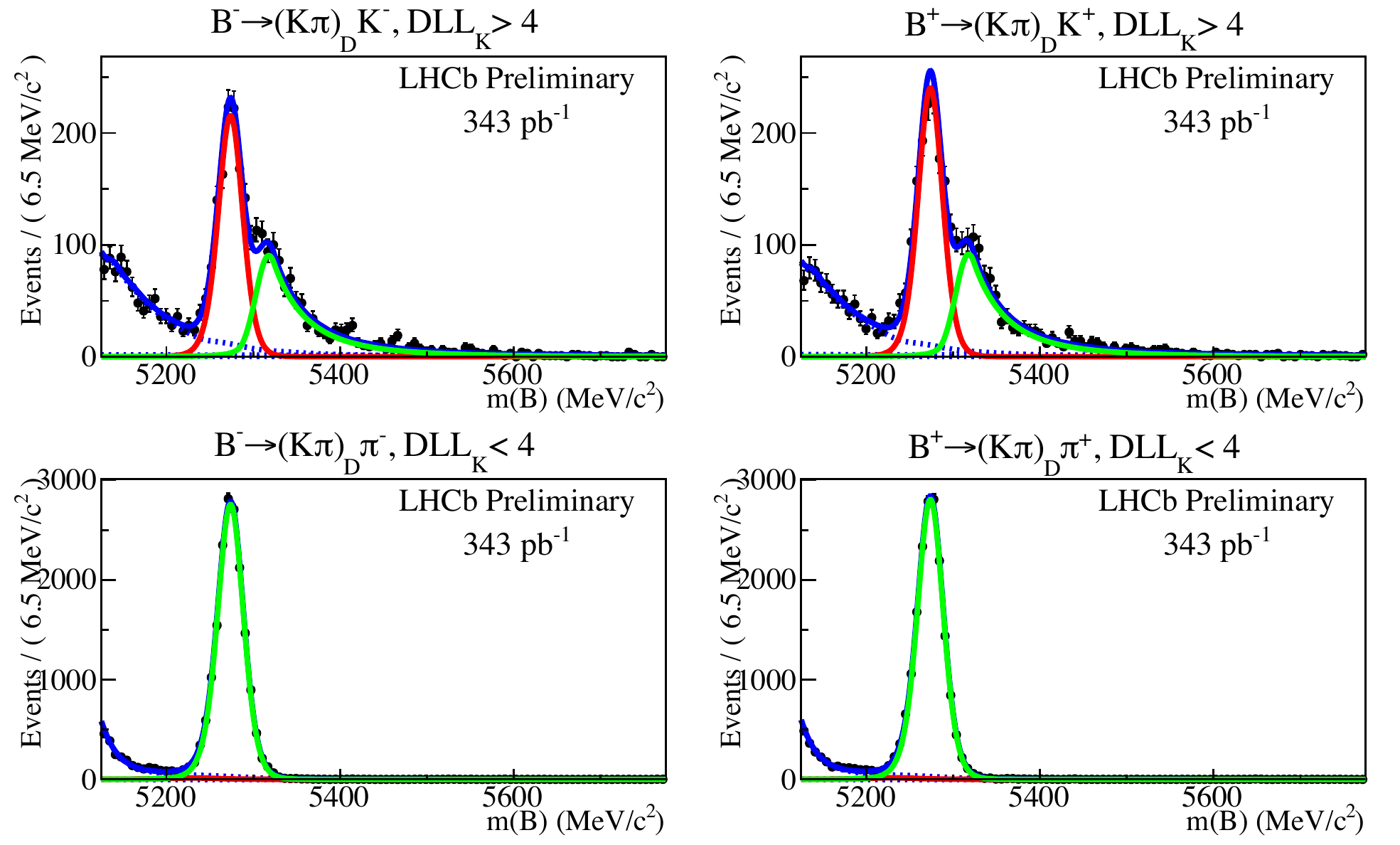}
    \caption{
      Signals for $B^\pm \to DK^\pm$, $D \to K^\pm\pi^\mp$ decays from LHCb~\cite{LHCb-CONF-2011-044}.
      (Top two plots) suppressed final states.
      (Bottom two plots) favoured final states.
      In each pair of plots the left (right) figure is for $B^-$ ($B^+$) decays.
      The plotted variable, $m(DK)$, peaks at the $B$ mass for signal decays, while background from $B^\pm \to D\pi^\pm$ appears as a satellite peak at positive values.
    }
    \label{fig:LHCb-DKADS}
  \end{center}
\end{figure}

Another powerful approach to constrain $\gamma$, the so-called ADS method~\cite{Atwood:1996ci,Atwood:2000ck}, comes from the use of doubly-Cabibbo-suppressed $D$ decays (for example to the final state $K^+\pi^-$).
Recent new results come from BaBar~\cite{delAmoSanchez:2010dz}, Belle~\cite{:2011tda} and CDF~\cite{Aaltonen:2011uu}, while the very latest results from LHCb~\cite{LHCb-CONF-2011-044} are shown in Fig.~\ref{fig:LHCb-DKADS}.
The world average for the parameter $R_{\rm ADS}$, which is the ratio of decay rates to the suppressed states compared to those for the favoured channels, including all these new results and illustrated in Fig.~\ref{fig:HFAG-DK} (right), shows that the suppressed decay is now clearly established, though no single measurement exceeds $5\,\sigma$ significance.
This is very promising for future $\gamma$ determinations.

Although the analyses with $B^\pm \to DK^\pm$ decays give the most precise results, different $B$ decays have also been studied.
The use of both possible decays $D^* \to D\pi^0$ and $D^*\to D\gamma$ provides an extra handle on the extraction of $\gamma$ from $B^\pm \to D^*K^\pm$~\cite{Bondar:2004bi} that is becoming visible in the most recent results~\cite{delAmoSanchez:2010dz,Belle-DKGLW}.
In addition, the $B_d^0 \to DK^{*0}$ channel may provide excellent sensitivity to $\gamma$, once sufficient statistics become available~\cite{Dunietz:1991yd,Gronau:2002mu,Gershon:2008pe,Gershon:2009qc}.

Until now, the strongest constraints on $\gamma$ from $B^\pm \to DK^\pm$ decays have come from analyses based on multibody $D$ decays, particularly $D \to K_S^0\pi^+\pi^-$, the so-called GGSZ method~\cite{Giri:2003ty}.
The most recent results\footnote{
  The BaBar measurement includes $B^\pm$ decays to $DK^\pm$, $D^*K^\pm$ and $DK^{*\pm}$ and $D$ decays to both $K_S^0\pi^+\pi^-$ and $K_S^0K^+K^-$, while the Belle results uses $DK^\pm$ and $D^*K^\pm$ with $D \to K_S^0\pi^+\pi^-$.
} are~\cite{delAmoSanchez:2010rq,Poluektov:2010wz}
\begin{equation}
  \gamma({\rm BaBar}) = \left( 68 \,^{+15}_{-14} \pm 4 \pm 3 \right)^\circ \, ,
  \hspace{5mm}
  \gamma({\rm Belle}) = \left( 78 \,^{+11}_{-12} \pm 4 \pm 9 \right)^\circ \, ,
\end{equation}
where the sources of uncertainty are statistical, systematic and due to imperfect knowledge of the amplitude model to describe $D \to K_S^0\pi^+\pi^-$ decays.
The last source can be eliminated by binning the Dalitz plot~\cite{Giri:2003ty,Bondar:2005ki,Bondar:2008hh}, using information on the average strong phase difference between $D^0$ and $\bar{D}^0$ decays in each bin that can be determined using quantum correlated $\psi(3770) \to D\bar{D}$ data samples.
The necessary measurements from $\psi(3770)$ data have recently been published by CLEO-c~\cite{Libby:2010nu} and used by Belle to obtain a model-independent result~\cite{:2011vfa}
\begin{equation}
  \gamma = \left( 77 \pm 15 \pm 4 \pm 4 \right)^\circ \, ,
\end{equation}
where the last uncertainty is due to the statistical precision of the CLEO-c results.  (Note that there is a strong statistical overlap in the data samples used for this result and the model-dependent Belle result reported above.\footnote{
  The Belle model-independent result uses only $B^\pm \to DK^\pm$ with $D \to K_S^0\pi^+\pi^-$.
})

Combining all available information on tree-level processes sensitive to $\gamma$, a world average can be obtained.
The values obtained by two different fitting groups are~\cite{Charles:2004jd,Bona:2005vz}
\begin{equation}
  \gamma ({\rm CKMfitter}) = \left(68 \, ^{+10}_{-11} \right)^\circ \, ,
  \hspace{5mm}
  \gamma ({\rm UTfit}) = \left( 76 \pm 9 \right)^\circ \, .
\end{equation}
The precision of the results is now reaching a level that the details of the statistical approach used to obtain the average value no longer has a strong influence on the uncertainty, but some difference in the central values is apparent.

An alternative approach to measuring $\gamma$, still using tree-level $B$ decays, is based on the time-dependent decay rates of $B_s \to D_s^\pm K^\mp$ processes~\cite{Aleksan:1991nh}.
This decay was previously observed by CDF~\cite{:2008rw}, and LHCb have recently reported clean signals~\cite{LHCb-CONF-2011-057}, that indicate that this mode can indeed be used to provide a competitive measurement of $\gamma$. 

It is also interesting to compare to the value of $\gamma$ obtained from processes that involve loop diagrams, since these may be affected by contributions from virtual, non-standard particles.
Recent results in charmless two-body $B$ decays are discussed in Refs.~\cite{kwon, raven}.
An interesting development in the last few years has been increased activity in the study of Dalitz plot distributions of charmless three-body $B$ decays, which can yield more information about the contributing amplitudes and hence can yield determinations of $\gamma$ that rely less strongly on theoretical input.
Results on $B_d^0 \to K_S^0 \pi^+\pi^-$~\cite{:2008wwa,Aubert:2009me} and $B_d^0 \to K^+\pi^-\pi^0$~\cite{:2011cxa} decays can be combined to provide a constraint on $\gamma$~\cite{Ciuchini:2006kv,Gronau:2007vr}.
The current data do not, however, provide a competitive result.

\subsection{Global CKM fits}

\begin{figure}[ht]
  \begin{center}
    \includegraphics[width=0.45\columnwidth]{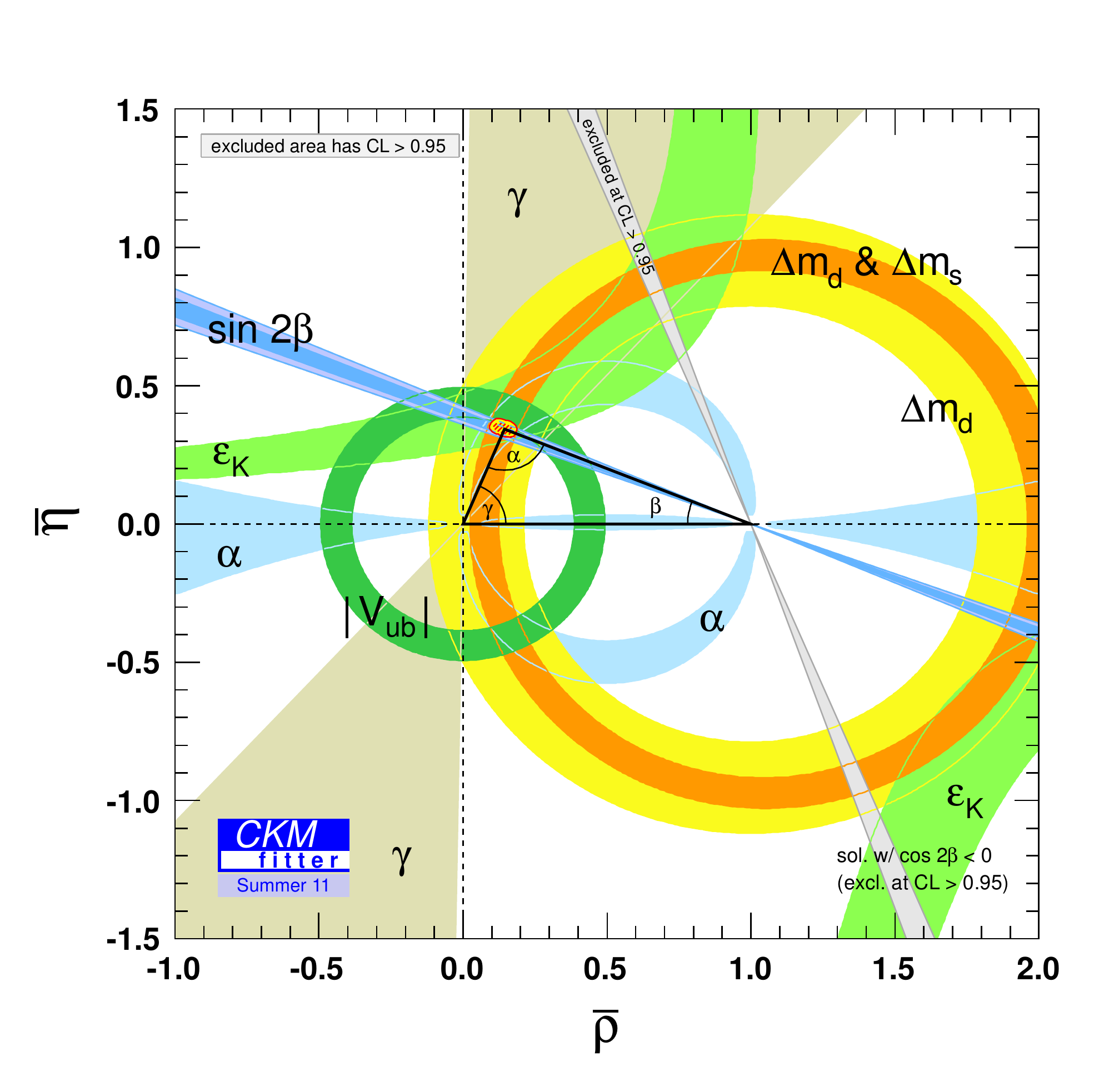}
    \includegraphics[width=0.45\columnwidth]{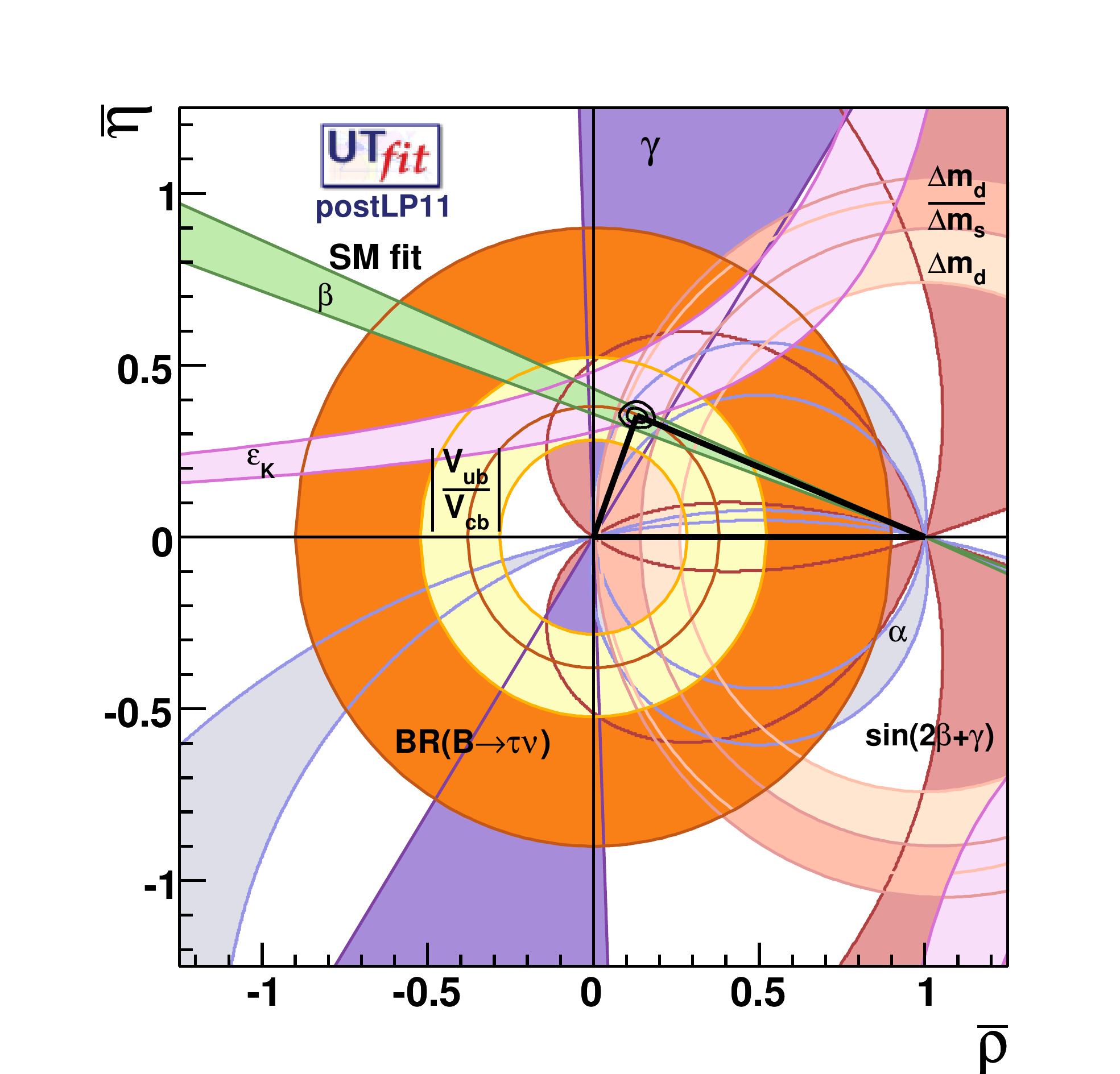}
    \caption{
      Results of global fits to the CKM Unitarity Triangle parameters, from (left) CKMfitter~\cite{Charles:2004jd} and (right) UTfit~\cite{Bona:2005vz}.
    }
    \label{fig:CKMfits}
  \end{center}
\end{figure}

The results of global fits to the CKM Unitarity Triangle parameters from the two main fitting groups, CKMfitter~\cite{Charles:2004jd} and UTfit~\cite{Bona:2005vz} are shown in Fig.~\ref{fig:CKMfits}.
The results are
\begin{eqnarray}
  \bar{\rho} = & 0.144 \, ^{+0.027}_{-0.018} \ ({\rm CKMfitter}) & = 0.132 \pm 0.020 \ ({\rm UTfit}) \, , \\
  \bar{\eta} = & 0.343 \pm 0.014 \ ({\rm CKMfitter}) & = 0.353 \pm 0.014 \ ({\rm UTfit}) \, .
\end{eqnarray}
In spite of the different statistical approaches used, consistent results are obtained.
The overall consistency of the different constraints on the ($\bar{\rho}$--$\bar{\eta}$) plane is good, though some tension exists (see, for example, Ref.~\cite{Lunghi:2010gv}).

\section{\boldmath Future prospects and conclusions}

The current time is certainly one of a changing era in quark flavour physics.
The previous generation of experiments is completing their analyses on their final data sets, while the next generation are commencing their programmes.
Looking further ahead, there are exciting prospects for this field of research, as new experiments are planned.
There is a future programme for kaon physics in the USA~\cite{Oddone}, and new experiments studying different aspects of kaon decays are also planned in Europe and Asia, among which the KLOE-2 experiment~\cite{AmelinoCamelia:2010me} has notable prospects to improve the measurement of $\left| V_{us} \right|$.

A new generation of high luminosity $e^+e^-$ machines, operating as $B$ factories or at lower energies is planned~\cite{Aihara,Browder:2008em}.
Another important step forward will occur with the upgrade of the LHCb detector~\cite{LHCb:1333091}, which will allow to exploit fully the flavour physics capability of the LHC.
The progress in theory must of course be matched by improved theoretical understanding.  
While such advances are in general hard to predict, there is very good reason to be optimistic that lattice calculations will continue to become more precise~\cite{lubicz}.

In conclusion, the CKM paradigm continues its unreasonable success, and despite some notable tensions with the SM, there is no discrepancy that can be considered proof of non-standard contributions.\footnote{
  At Lepton Photon 2011, the author compared the long wait to discover effects beyond the SM to that for Indian batting hero Sachin Tendulkar to achieve his $100^{\rm th}$ century in international cricket.  
  Sadly, at the time of writing these proceedings, and despite some close calls, we are still waiting for both historic achievements.
}
Nonetheless, there is reason for optimism, since current and future projects promise significant improvements.
In the short term, the BESIII and LHCb experiments, together with improved lattice calculations will, at the very least, advance our knowledge and may provide a breakthrough.
In the certainty that new sources of $CP$ violation exist, somewhere, and with various other reasons to expect non-SM physics around the TeV scale (or higher) to cause observable effects in flavour-changing interactions in the quark sector, continued study of the elements of the CKM matrix remains a key cornerstone of the global particle physics programme.

\acknowledgments
I am grateful to help from individuals from the BaBar, Belle, CLEO-c, KLOE, LHCb and NA62 experiments, the working group conveners from CKM2010 and the other speakers in the Flavour Physics sessions at Lepton Photon 2011.
I would particularly like to thank Marcella Bona, Erika De Lucia, Vera Luth, Karim Trabelsi and Guy Wilkinson.
This work was supported by the EU under FP7.

\bibliographystyle{pramana}
\bibliography{references}

\end{document}